\documentclass[aps,prd,showpacs]{revtex4}
\usepackage{graphicx}
\usepackage{psfig}
\usepackage{revsymb}
\begin{document}
\renewcommand{\thesection}{\arabic{section}}
\renewcommand{\thesubsection}{\arabic{subsection}}
\title{The Study of Relatively Low Density Stellar Matter in Presence of Strong
Quantizing Magnetic Field}
\author{Nandini Nag, Sutapa Ghosh and Somenath 
Chakrabarty$^\dagger$}
\affiliation{
Department of Physics, Visva-Bharati, Santiniketan 731 235, 
West Bengal, India\\ 
$^\ddagger$E-mail:somenath.chakrabarty@visva-bharati.ac.in}
\begin{abstract}
The effect of strong quantizing magnetic field on the equation of state of 
matter at the outer crust region of magnetars is studied. The density of such 
matter is low enough compared to the matter density at the inner crust or outer
core region. Based on the relativistic version of semi-classical
Thomas-Fermi-Dirac model in presence of strong quantizing magnetic field
a formalism is developed to investigate this specific problem. The equation of 
state of such low density crustal matter is obtained by replacing the 
compressed atoms/ions by Wigner-Seitz cells with nonuniform electron
density. The results are compared with other possible scenarios. The appearance
of Thomas-Fermi induced electric charge within each Wigner-Seitz cell is also 
discussed. 
\end{abstract}
\maketitle
\section{Introduction}
The theoretical investigation on the properties of compact stellar objects in 
presence of strong quantizing magnetic field have gotten a new life after the 
discovery of a large number of magnetars \cite{R1,R2,R3,R4}. These exotic 
stellar objects are believed to be strongly magnetized young neutron stars. 
Their surface magnetic fields are observed to be $\geq 10^{15}$G. Then it is 
quite possible that the fields at the core region may go up to $10^{18}$G (the 
theoretical estimate may easily be obtained from scalar virial theorem). The 
exact source of such strong magnetic field is of course yet to be known. These 
objects are also supposed to be the possible sources of anomalous X-ray and 
soft gamma emissions (AXP and SGR). Now, if the magnetic field is really so 
strong, specially at the core region, it must affect most of the important 
physical properties of these stellar objects and also some of the physical
processes, e.g., the rates / cross-sections of elementary processes. In 
particular the weak and the electromagnetic decays / reactions taking 
place at the core region will change significantly.

The strong magnetic field affects the equation of state of dense neutron
star matter. As a consequence the gross-properties of neutron stars 
\cite{R5,R6,R7,R8}, e.g., mass-radius relation, moment of inertia,
rotational frequency etc. should change significantly. In some recent work
we have developed the relativistic version of Landau theory of Fermi liquid
in presence of strong quantizing magnetic field to obtain the equation of
state of dense neutron star matter \cite{SCANN,SCLAND}. We have also shown
that the nucleons acquire complex mass in the $\sigma -\omega -\rho$ meson 
exchange type mean field in presence of strong magnetic field. It has
been noticed that due to the complex nature of neutron and proton
energies, there is a kind of relaxation or oscillation in the iso-spin
space. In the case of compact neutron stars, the phase transition from neutron 
matter to quark matter, which may occur at the core region, is also affected by
strong quantizing magnetic field. It has been shown that a first order phase 
transition initiated by the nucleation of quark matter droplets is absolutely 
forbidden if the magnetic field strength $\sim 10^{15}$G at the core region 
\cite{R9,R10}. However, a second order phase transition is allowed, provided 
the magnetic field strength $<10^{20}$G. This is of course too high to achieve 
at the core region. The study of time evolution of nascent quark matter, 
produced at the core region through some higher order phase transition, shows 
that in presence of strong magnetic field it is absolutely impossible to
achieve chemical equilibrium ($\beta$-equilibrium) configuration among
the constituents of the quark phase if the magnetic field strength is as
low as $ B \sim 10^{14}G$.

The elementary processes, in particular, the weak and the electromagnetic
decays/reactions taking place at the core region of a neutron star are strongly 
affected by such ultra-strong magnetic fields \cite{R11,R12}. Since the cooling 
of neutron stars are mainly controlled by neutrino/anti-neutrino emissions, the 
presence of strong quantizing magnetic field should affect the thermal history 
of strongly magnetized neutron stars. Further, the electrical conductivity of 
neutron star matter, which mainly comes from free electron gas within the stars 
and directly controlls the evolution of neutron star magnetic field, should 
also change significantly \cite{R12}.

In another kind of work, the structural stability of such strongly magnetized 
rotating objects are studied. It has been observed from the detailed general 
relativistic calculation that there are possibility of some form of geometrical
deformation of these strongly magnetized objects from their usual spherical 
shapes \cite{RR13,RR14,RR15}. In the extreme case such objects may either 
become black strings or black disks. It is quite possible to have gravity wave 
emission from these rotating magnetically deformed objects (see also 
\cite{SOMA} for the magnetic deformation of nucleonic bags in neutron star 
matter).

In a recent study of the microscopic model of dense neutron star matter,
it has been observed that if most of the electrons occupy
the zeroth Landau level, with spin anti-parallel to the direction of
magnetic field, and only a few are with spin along the direction of
magnetic field  with non-zero Landau quantum number, then either such strongly
magnetized system can not exist or such a strong magnetic field can not be 
realized at the core region of a neutron star \cite{RR16}. We have observed
that $10^{22}$G is the upper limit of magnetic field which the core of a
neutron star can sustain. We have also noticed that since the electrical 
conductivity of the medium becomes extremely low in presence of ultra-strong 
magnetic field, the magnetic field at the core region must therefore decay very 
rapidly. Hence it may be argued that the magnetic field at the core of a 
magnetar can not be too high, it is only the strong surface field which has 
been observed.

Similar to the study of quark-hadron deconfinement transition inside neutron 
star core in presence of strong quantizing magnetic field, a lot of 
investigations have also been done on the effect of ultra-strong magnetic 
field on chiral properties of dense quark matter \cite{SCCH,R13,R14,R15,R16,
R17,R18,R19,Rina,VP2,SPK,WO} (see also \cite{wang} for the chiral properties of
strongly magnetized electron gas). The strong external magnetic field acts like
a catalyst to generate mass dynamically. This is one of the most important 
effect of ultra-strong magnetic field on the properties of charged elementary 
particles. This effect is of course impossible to verify in the terrestrial 
laboratory. Only the cosmic laboratory offers us the opportunity to test this 
intrinsic phase transition in the world of elementary particles. Therefore, the
presence of strong magnetic field in stellar matter causes a lot of interesting
and significant changes in the properties of elementary particles. Also a lot of
new phenomena are observed in such cosmic laboratory in presence of ultra-strong
magnetic field.

The exotic process of magnetic photon splitting, i.e., the decay of a photon 
into two photons (or combination of two photons into a single one) in presence 
of a very strong magnetic field, has recently attracted renewed attention, 
mainly because of the great importance this process may have in the
interpretation of the spectra of cosmic $\gamma$-ray burst sources.
The basic formulae for magnetic photon splitting had already been
derived in the seventies \cite{ADLER} (see also \cite{ADLER1,WILKE,BAIER}).
Again, since such strong magnetic field can not be generated, even in the
pulse form, in the laboratory, it is absolutely impossible to verify this
exotic phenomenon of photon splitting.

In this article we shall develop an exact (within the limitation of
Thomas-Fermi-Dirac model) formalism of relativistic version of 
Thomas-Fermi-Dirac model in presence of strong quantizing magnetic field. From 
this model calculation, we shall obtain the equation of state of relatively low
density crustal matter of magnetars, which is mainly dense iron crystal. We 
shall replace the iron ions / atoms by Wigner-Seitz cells with non-uniform 
electron density within each cell. We have organized the paper in the following
manner: In the next section, we shall present the basic formalism of 
relativistic Thomas-Fermi-Dirac equation in presence of strong quantizing 
magnetic field. In section 3, we shall evaluate the equation of state of low 
density crustal matter of strongly magnetized neutron star. The kinetic energy 
part of the electron gas within each cell is obtained in section 4. Section 5 
is dealt with the electron-nucleus interaction energy. In section 6, we 
evaluate the electron-electron direct interaction energy part. The exchange 
interaction part is obtained in section 7. The possibility of Thomas-Fermi 
induced charge within each cell is discussed in section 8 and finally in 
section 9, the last section, we shall discuss the results and future prospects 
of this model. 

It is worth mentioning that the formalism we are presenting
in this article is also applicable to strongly magnetized white dwarfs. For
the sake of completeness, we have compared the results of the present model,
with other possible physical scenarios, e.g., non-relativistic case with zero
and non-zero values of Landau quantum number, with the non-relativistic
and relativistic field free cases, etc.

Finally, we would like to mention, that the properties of low density magnetized
crustal matter, mainly the electromagnetic properties, which also includes the 
transport properties have been studied thoroughly by Potekhin and Potekhin et. 
al. \cite{RU}. 

\section{Relativistic Thomas-Fermi-Dirac Model in Quantizing Magnetic Field}
In the past few decades, a lot of work have been done on both
Thomas-Fermi and Thomas-Fermi-Dirac models in absence as well as in
presence of strong magnetic fields to obtain the equation of
state of low density crustal matter of compact stellar objects. Unfortunately, 
the calculations are either based on some crude approximation or are incomplete
in nature. In those calculations, in presence of strong quantizing magnetic 
field, the occupancy of only the zeroth Landau level by electrons were 
considered. However, such approximation is valid if the magnetic fields are 
extremely strong. In one of the previous calculations we also made a 
preliminary study of low density crustal matter of neutron stars in presence of
strong quantizing magnetic field using the Thomas-Fermi-Dirac approach. The 
model was non-relativistic in nature and occupancy of only the zeroth Landau 
level was taken into account \cite{NAG}. In some work, more than a decade ago, 
Shivamogi and Mulser did some relativistic calculation for atoms in 
ultra-strong magnetic field \cite{SHIV}. Again, the calculation is 
applicable for the system where only the zeroth Landau level is occupied. A 
similar type of calculation was also done without magnetic field by Ruffini 
\cite{REMO}. At this point we should mention that Thomas-Fermi model is 
not an exact method. It is a semi-classical approach. It was shown by Lieb
and Simon that this model will be an exact one for atoms, molecules or 
solid in general, if the atomic number $Z\rightarrow \infty$ \cite{LIEB}
(see also \cite{LIEB1} and \cite{LIEB2} for the very nice review articles).
The Thomas-Fermi model in presence of ultra-strong magnetic field (when only 
the zeroth Landau level is populated) in the non-relativistic
regime was first used by Kadomtsev \cite{K1}. It was shown that in presence
of strong magnetic field, the electrons move in cylindrical shells with the
axis directed along the magnetic field. The atoms thus have elongated
cylindrical shapes and much more binding energies. The Thomas-Fermi model
for heavier atoms have also been studied by Mueller et. al., using a
variational approach \cite{MU12}. The relativistic corrections of these
calculations was done by Hill et. al. \cite{HILL}. But in each of these
studies an extremely strong magnetic field was considered, so that electrons
occupy only the zeroth Landau level.

To develop an exact formalism for relativistic Thomas-Fermi-Dirac model in
presence of strong quantizing magnetic field at zero temperature (it is
assumed that the electron gas within the cell is strongly degenerate), we 
assume that the magnetic field $\vec B$ is uniform throughout the star and is 
along $Z$-direction, i.e., our choice of gauge is $A^\mu\equiv(0,0,xB,0)$.
Now in the relativistic scenario, the Landau levels of the electrons 
will be populated if the magnetic field strength $B$ exceeds 
the quantum critical value $B^{(c)(e)}=m^2/e \approx 4.4\times 10^{13}$G
(throughout this article we assume $\hbar=c=1$). In the relativistic regime the
quantum critical value is the typical strength of the magnetic field at which 
the electron cyclotron quantum exceeds the corresponding rest mass energy or 
equivalently the de Broglie wave length for electron exceeds its Larmor radius. 
In presence of such strong quantizing magnetic field along $Z$-axis the 
electron momenta in the orthogonal plane get quantized and are given by 
$p_\perp=(2\nu eB)^{1/2}$, where $\nu=0,1,2 ....$, the well known Landau 
quantum numbers. The component along $Z$-axis varies continuously from 
$-\infty$ to $\infty$, for non-zero temperature, whereas, in the zero 
temperature case, we have the relation: $-p_F \leq p_z \leq +p_F$, where $p_F$ 
is the electron Fermi momentum. The phenomenon is known as the Landau 
quantization. Further, the phase space volume integral in the momentum space 
in this quantized condition is given by
\begin{equation}
\frac{1}{(2\pi)^3}\int d^3p f(p)=
\frac{1}{(2\pi)^3}\int dp_z d^2p_\perp
f(p)=\frac{eB}{4\pi^2}\sum_{\nu=0}^{\nu=\infty}
(2-\delta_{\nu 0})\int_{-\infty}^{+\infty} dp_z f(\nu,p_z)
\end{equation}
The presence of the multiplicative factor $2-\delta_{\nu 0}$ is justified by 
the fact that the zeroth Landau level is singly degenerate, whereas all other
states are doubly degenerate (which will only be obvious if one solves
Dirac equation in presence of strong external magnetic field of strength $B
> B^{(c)(e)}$). The modified form of spinor solutions of Dirac equation for 
electrons in Dirac-Pauli representation, in presence of strong quantizing 
magnetic fields are given by
\begin{equation}
\psi(x)=\frac{1}{(L_yL_z)^{1/2}}\exp\{-iE_\nu t+ip_yy+ip_zz\}u^{\uparrow
\downarrow}(x)
\end{equation}
where
\begin{equation}
u^\uparrow(x)=\frac{1}{[2E_\nu(E_\nu+m)]^{1/2}} \left (\begin{array}{c}
(E_\nu+m)I_{\nu;p_y}(x)\\ 0\\ p_zI_{\nu;p_y}(x)\\ -i(2\nu eB)^{1/2}
I_{\nu-1;p_y}(x) \end{array} \right )
\end{equation}
and
\begin{equation}
u^\downarrow(x)=\frac{1}{[2E_\nu(E_\nu+m)]^{1/2}} \left (\begin{array}{c}
0\\ (E_\nu+m)I_{\nu-1;p_y}(x)\\ i(2\nu eB)^{1/2} I_{\nu;p_y}(x)\\
-p_zI_{\nu-1;p_y}(x)
\end{array} \right )
\end{equation}
where the symbols $\uparrow$ and $\downarrow$ indicates up and down spin states 
respectively,
\begin{equation}
I_\nu=\left (\frac{qB}{\pi}\right )^{1/4}\frac{1}{(\nu !)^{1/2}}2^{-\nu/2}
\exp \left [{-\frac{1}{2}eB\left (x-\frac{p_y}{eB} \right )^2}\right ]
H_\nu \left [(eB)^{1/2}\left (x-\frac{p_y}{eB} \right) \right ]
\end{equation}
with $H_\nu$ is the well known Hermite polynomial of order $\nu$, $E_\nu= 
(p_z^2+m^2+2\nu eB)^{1/2}$, the single particle energy eigen value, $L_y$, $L_z$
are the length scales along $Y$ and $Z$ directions respectively, $e$ is the 
magnitude of the charge carried by electrons and $m$ is the electron rest mass.

Now for $B >B^{(c)(e)}$, using the inequality $p_F^2 \geq 0$, the upper limit 
of Landau quantum number upto which can be occupied at zero temperature is 
given by
\begin{eqnarray}
\nu_{\rm{max}}=\left [\frac{(\mu_{e}^{2}-m^{2})}{2eB}\right ]
\end{eqnarray} 
which is an integer but less than the actual value of the quantity within the
third brackets at the right hand side and $\mu_e$ is the electron chemical 
potential. At zero temperature, the upper limit of $\nu$-sum will be 
$\nu_{\rm{max}}$ instead of $\infty$. The external magnetic field will 
therefore behave like a classical entity if the strength is less than the 
quantum threshold value and in this region one has to take the standard form of
plane wave solution of Dirac spinors. Whereas, in presence of ultra-strong 
magnetic field, the maximum value of the Landau quantum number as mentioned 
above becomes zero. Which indicates that all the electrons occupy the zeroth 
Landau level have their spins aligned opposite to the direction of magnetic 
field. Following eqn.(1), it is very easy to show that in presence of 
strong quantizing magnetic field the number density for electron is given by
\begin{equation}
n_e=\frac{eB}{2\pi^2}\sum_{\nu=0}^{\nu_{\rm{max}}}(2-\delta_{\nu 0}) p_F
\end{equation}
In the Thomas-Fermi-Dirac model in presence of strong magnetic field in the
relativistic region, we replace the atoms/ions by Wigner-Seitz cells with
varying electron density and assume that $A$ and $Z$ are the mass number and 
atomic number within the cell ($A=N+Z$, with $N$, the number of neutrons). To 
make each cell charge neutral, $Z$ must also be the number of electrons within 
the cell. We further assume that instead of a point object, the radius of the 
nucleus is given by $r_n=r_0A^{1/3}$ with $r_0=1.12$fm. We will see later that 
this choice will remove the singularity of Thomas-Fermi equation at the origin. 
The electrostatic potential $V(r)$, felt by electrons satisfy the Poisson's 
equation, given by
\begin{equation}
\nabla^2 V(r)=4\pi e n_e(r)-4\pi en_p(r)
\end{equation}
where $n_p(r)$ is the proton density within the nucleus, given by
\begin{equation}
n_p=\frac{3Ze}{4\pi r_n^3}\theta(r_n-r)
\end{equation}
The second term on the right hand side of eqn.(8) is nuclear contribution.
Now the maximum energy of an electron at $\vec r$ within the cell is given by
\begin{eqnarray}
\varepsilon_\nu(r)- eV(r)&=&{\rm{constant}}=\mu_e\\
~~ {\rm{or}} ~~~~~~
(p_F^2+m^2+2\nu eB)^{1/2} - eV(r)&=&{\rm{constant}}=\mu_e
\end{eqnarray}
In case it is not a constant, the electrons will try to occupy a position
within the cell to have minimum energy. This will develop an instability in the
system. From the above equation we have
\begin{equation}
p_F=[(\mu_e+eV(r))^2-m_\nu^2]^{1/2}
\end{equation}
where $m_\nu^2=m^2+2\nu eB$. Since we are interested to have electron
distribution only outside the nucleus, we discard the proton contribution in
the Poisson's equation. Further, the potential must satisfy the following
boundary conditions:
\begin{eqnarray}
&& rV(r)=Ze ~~~~~~~{\rm{for}}~~~~ r\rightarrow r_n ~~~~~~~(a)\nonumber\\
&& \frac{dV}{dr}=0 ~~~~~~~~~~~~{\rm{for}}~~~~ r\rightarrow r_s ~~~~~~~~(b)
\nonumber
\end{eqnarray}
where $r_s$ is the surface value of $r$ (the geometrical structure of each cell 
is assumed to be spherical in nature). The Poisson's equation is then given by
\begin{equation}
\frac{1}{r}\frac{d^2}{dr^2}(rV(r))=\frac{2e^2B}{\pi}
\sum_{\nu=0}^{\nu_{\rm{max}}}(2-\delta_{\nu
0})[(\mu_e+eV(r))^2-m_\nu^2]^{1/2}
\end{equation}
We now substitute
\begin{eqnarray}
&& \mu_e+eV(r)=Ze^2\frac{\phi(r)}{r} ~~~~~~~~(c)\nonumber \\
&& r=\mu x ~~~~~~~~~~~~~~~~~~~~~~~~~~~~~ (d)\nonumber
\end{eqnarray}
Then the modified form of the Poisson's equation is given by
\begin{equation}
\frac{d^2\phi}{dx^2}=\sum_{\nu=0}^{\nu_{\rm{max}}}(2-\delta_{\nu 0}) (\phi^2(x)-
\phi_0^2 x^2)^{1/2}
\end{equation}
with 
\[
\mu=\left ( \frac{\pi}{2e^3B}\right )^{1/2} ~~~~{\rm{and}}~~~~ \phi_0= 
\frac{m_\nu \mu}{Z e^2} 
\]
From (c) and (d), it is therefore quite obvious, that the radius of each
spherical cell, at the crustal region, decreases with the strength of
magnetic field and the change is $\propto B^{-1/2}$. The squeezing of the 
Wigner-Seitz 
cells in presence of strong magnetic field is analogous to the well known 
magneto striction phenomenon observed in classical magneto-static problem.
Further, the right hand side of the final form of the Poisson's equation (eqn.(14))
must be real. It requires, that the inequality $\phi_0 x \leq
\phi(x)$ must be satisfied. Which after using $\phi_0$ and $m_\nu$, gives
\begin{equation}
\nu_{\rm{max}}(x)=\left [ \frac{e^4Z^2}{\pi
x^2}\phi^2(x)-\frac{m^2}{2eB}\right ] \geq 0
\end{equation}
This equation indicates that the upper limit of the Landau quantum number of
the levels occupied by the Wigner-Seitz electrons depends on its position 
within the cells. Or in other wards, the value of $\nu_{\rm{max}}$ at some 
point $x$ within the cell also depends on the strength of electrostatic 
potential at that point (of course it is assumed that the magnetic field is 
constant throughout the star). Since we have assumed a finite dimension for 
the nucleus, the problem of singularity at the origin will not appear here 
(origin is actually excluded in our numerical calculation, see also 
\cite{REMO}). Since the minimum value of $\nu_{\rm{max}}$ is 
zero for ultra-strong magnetic field case, we have a maximum value of atomic 
radius, given by $r_{\rm{max}}=\mu x_{\rm{max}}$, where
\begin{equation}
x_{\rm{max}}\leq \left ( \frac{2eB}{\pi} \right )^{1/2} \frac{e^3 Z}{m}
\phi(x_{\rm{max}})
\end{equation}
The value of $x_{\rm{max}}$, obtained from the numerical solution of
the Poisson's equation, must satisfy this inequality. Further, it is very easy to 
show from the conditions (a) and (b), using (c) and (d), that the initial and 
the surface condition for $V(r)$ are given by
\begin{equation}
\phi(x)\vert_{x=x_n}=1 ~~{\rm{and}}~~\frac{d\phi(x)}{dx}\vert_{x=x_s}=
\frac{\phi(x)}{x}\Big \vert_{x=x_s},
\end{equation}
respectively. These conditions are identical with the field free case
\cite{WO}. While 
solving the Poisson's equation numerically using the two boundary conditions 
(a) and (b) (In the 
numerical calculation we have followed the standard 4-point Runge-Kutta  
technique with shooting method at the surface), we found that it is absolutely 
necessary to incorporate all other extra conditions appearing in this 
particular case (eqns.(15) and (16)). The surface condition must also satisfy
\[
Z=4\pi \int_{r_n}^{r_s}r^2 n_e(r)dr=4\pi\mu^3\int_{x_n}^{x_s}x^2n_e(x) dx
\]
Since the integration range is from $r_n$ to $r_s$ (origin is avoided), the
serious problem of singularity associated with Thomas-Fermi-Dirac equation, 
at the centre of the cell will no longer be there. Therefore, in this 
particular case, it is also not necessary to follow the 
numerical methods prescribed by Feynman, Metropolis and Teller \cite{FEY}.

The corresponding expression for the Poisson's equation for $B=0$ in the 
non-relativistic regime is given by \cite{SHA} (see also \cite{WO} and \cite{MER})
\[
\frac{d^2\phi}{dx^2}=\frac{\phi^{3/2}}{x^{1/2}}
\]
where 
\[
\phi(x=0)=1 ~~~{\rm{and}}~~~ \phi^\prime(x_s)=\frac{\phi(x_s)}{x_s}
~~~{\rm{and}}~~~ \mu=\left ( \frac{9\pi^2}{128Z}\right )^{1/3}a_0
\]
with $a_0=\hbar^2/(me^2)$, the Bohr radius. This equation has a
singularity at the origin.

To obtain the variation of $\phi(x)$ with $x$ within a typical
Wigner-Seitz cell, we have solved the Poisson's equation (eqn.(14))
numerically considering all the necessary conditions as mentioned above
along with the boundary conditions (a) and (b). In fig.(1) we have plotted 
$\phi(r)$ as a function of $r$ (in \AA),~ the radius of the cell, for
three different initial values for $\phi^\prime$ and magnetic field strength
$B=10^{14}$G. In fig.(2) we have plotted the same quantity but for three
different magnetic field strengths. It is quite obvious from the curves
of fig.(2) that the surface of the cell is reached later for low
magnetic field strength compared to stronger values. This is consistent
with the well known magneto striction phenomenon.

\section{Equation of state}
To obtain the equation of state of such low density crustal matter in
presence of strong quantizing magnetic field in the relativistic region,
we first evaluate the kinetic pressure, given by 
\begin{equation}
P=\frac{eB}{2\pi^2}\sum_{\nu=0}^{\nu_{\rm{max}}} (2-\delta_{\nu 0})
\int_0^{p_F}\frac{p_z^2} {(p_z^2+m_\nu^2)^{1/2}} dp_z
\end{equation}
This momentum integral can very easily be evaluate and is given by 
\begin{equation}
P=\frac{eB}{2\pi^2}\sum_{\nu=0}^{\nu_{\rm{max}}}(2-\delta_{\nu 0}) \left [
p_F(p_F^2+m_\nu^2)^{1/2}- m_\nu^2 \ln \left (\frac{p_F+(p_F^2+m_\nu^2)^{1/2}}
{m_\nu} \right ) \right ]
\end{equation}
where $p_F=p_F(x_s)$. The above expression, therefore gives the cell averaged 
kinetic pressure. To obtain the numerical values, it is necessary to express 
$p_F(x_s)$ as a function of $\phi(x_s)$ using eqn.(21) as given below (either 
the correct expression, particularly for numerical evaluation or the
approximate one for the analytical result). For the exact expression,
as given in eqn.(19), the functional relation is extremely complicated and 
has to be obtained numerically. To get a simple mathematical expression for 
kinetic pressure, we consider the ultra-relativistic case, which gives
\begin{equation}
P=\frac{eB}{4\pi^2}\sum_{\nu=0}^{\nu_{\rm{max}}}(2-\delta_{\nu 0})
p_F^2(x_s)
\end{equation}
Now to express the Fermi momentum as a function of $\phi(x_s)$ in the 
ultra-relativistic limit, we consider
\[
p_F=[(\mu_e+eV(r))^2-m_\nu^2]^{1/2} =\left [ Z^2e^4 \left (
\frac{\phi(x_s)}{\mu x_s}\right )^2-m_\nu^2 \right ]^{1/2}, 
\]
which is the correct expression. On putting this correct expression for
Fermi momentum in eqn.(19), the exact functional dependence of kinetic
pressure on $\phi(x_s)$ can be obtained. The approximate form may be obtained 
by neglecting $m_\nu$, and we have
\[
p_F\approx  \mu_e+eV(r)
\]
Substituting the conditions (c) and (d), we have
\begin{equation}
p_F\approx Ze^2\frac{\phi(r_s)}{r_s}=Ze^2 \frac{\phi(x_s)}{\mu x_s}
\end{equation}
Hence
\begin{equation}
P\approx \frac{eB}{4\pi^2}\sum_{\nu=0}^{\nu_{\rm{max}}} (2-\delta_{\nu 0})
Z^2e^4 \left (\frac{\phi(x_s)}{\mu x_s}\right )^2
\end{equation}
In this approximation, the electron number density is given by 
\begin{equation}
n_e=\frac{eB}{2\pi^2}\sum_{\nu=0}^{\nu_{\rm{max}}} Ze^2 \frac{\phi(x_s)}
{\mu x_s}
\end{equation}
Whereas, the usual form of mass density or rest energy density 
of the crustal region of neutron stars, contributed by the
massive nuclear part, is given by  
\begin{equation}
\rho(x_s)=\epsilon(x_s)=\frac{3Am_B}{4\pi \mu^3 x_s^3}
\end{equation}
Here, the nuclei are assumed to be static in nature and are not affected by 
strong magnetic field. Whereas, electrons within the cells are considered to 
be almost free and dominates the kinetic pressure part. On eliminating $x_s$ 
from eqn.(24) and either from eqn.(19) or eqn.(22), we get the equation of 
state of low density crustal matter in exact or approximate form respectively.

\section{Kinetic Energy for Electron Gas}
In this section we shall evaluate the energy contribution from the electron 
part within the cell. Although it is small enough compared to the rest energy 
from nuclear part in the non-relativistic region, we have noticed that it has 
a significant amount of contribution in the relativistic scenario. The
energy contribution from the electronic part consists of four main parts,
they are: (i) the kinetic energy part, (ii) the electron-nucleus interaction part,
(iii) the electron-electron direct interaction part and (iv) the electron-electron
exchange interaction part. There can be another important contribution, the
correlation part. Now the correlation energy is not a quantity with physical 
significance. It actually gives the error incurred in making a fairly crude 
approximation. In the Hartree zeroth order approximation, the $N$-electron 
distribution function factors into a product of $N$ one-electron distributions.
In the Hartree-Fock wave function (Slatter determinant form) correlations 
are introduced in the first order approximation. In the evaluation of actual 
correlation term, higher order approximations are incorporated and the 
exchange term is excluded. Since the exchange term is absorbed in the many 
body correlation terms, in the density functional theory, it is conventionally 
known as the exchange-correlation term and obtained with some approximation 
of electron density. In this article, however, we have not taken the 
correlation part into account. In some future communication we shall bring 
this extra term in the energy expression and evaluate numerically the 
equation of state of low density crustal matter.

Now the contribution from electrons to the kinetic energy part is given by
\begin{equation}
E_{KE}=\int_{r_n}^{r_s} d^3r \frac{eB}{2\pi^2}\sum_{\nu=0}^{\nu_{\rm{max}}}
(2-\delta_{\nu 0})\int_0^{p_F} dp_z [(p_z^2+m_\nu^2)^{1/2}-m]
\end{equation}
Evaluating the integral over $p_z$, the exact expression for kinetic
energy is given by
\begin{equation}
E_{KE}=\frac{eB}{\pi}
\int_{r_n}^{r_s}
\sum_{\nu=0}^{\nu_{\rm{max}}}(2-\delta_{\nu 0})
r^2dr \left [ p_F(p_F^2+m_\nu^2)^{1/2}+
m_\nu^2\ln\left (\frac{p_F+(p_F^2+m_\nu^2)^{1/2}}{m_\nu}\right ) -2mp_F
\right ] 
\end{equation}
where the exact form of $p_F$ is given by
\begin{eqnarray}
p_F(x)&=&\left [\frac{Z^2e^4\phi(r)^2}{r^2}-m_\nu^2\right ]^{1/2}\nonumber \\ 
&=&\left [ \frac{Z^2e^4\phi(x)^2}{\mu^2x^2}-m_\nu^2\right ]^{1/2}
\end{eqnarray}
Again this will give an extremely complicated functional dependence on
$\phi(x)$. In addition, since $\nu_{\rm{max}}(x)$ depends on the radial 
coordinate, the evaluation of $E_{KE}$ by analytical integration is just 
impossible. Therefore, numerical technique has to be followed to evaluate 
the integrals along with the sum over $\nu$. Numerically fitted functional 
forms for $\phi(x)$ and the corresponding $\nu_{\rm{max}}(\phi(x))$ are used 
to evaluate the kinetic energy numerically. To get a simpler expression, we go 
to the ultra-relativistic limit, then 
\begin{equation}
E_{KE}\approx \frac{eB\mu^3}{\pi} 
\int_{x_n}^{x_s}
\sum_{\nu=0}^{\nu_{\rm{max}}}(2-\delta_{\nu 0})
p_F^2(x)x^2dx
\end{equation}
To obtain an approximate result, we use the relations
\begin{equation}
p_F\approx Ze^2\frac{\phi(x)}{\mu x},
\end{equation}
and 
\begin{equation} 
\sum_{\nu=0}^{\nu_{\rm{max}}}(2-\delta_{\nu 0})\phi^2(x)\approx
\phi^{\prime\prime}(x) + \sum_{\nu=0}^{\nu_{\rm{max}}}(2-\delta_{\nu 0})
\phi_0^2x^2
\end{equation}
Then integrating by parts the Poisson's equation, considering the surface
condition and finally with the trivial relation (we further assume that
$\nu_{\rm{max}}$ is a function of $x_s$, instead of a function of $x$)
\[
\sum_{\nu=0}^{\nu_{\rm{max}}}(2-\delta_{\nu 0})\nu=\nu_{\rm{max}}
(\nu_{\rm{max}}+1),
\]
we have
\begin{equation}
E_{KE}=\frac{Z^2e^2}{2\mu}\left [ \left \{\frac{\phi(x_s)}{x_s}
-\phi^\prime(x_n)\right \} +\frac{\pi}{3Z^2e^6}(x_s^3-x_n^3)\nu_{\rm{max}}
(\nu_{\rm{max}}+1)\right ]
\end{equation}
For ultra-strong magnetic field, $\nu_{\rm{max}}=0$, then we have
\begin{equation}
E_{KE}=\frac{Z^2e^2}{2\mu}\left [ \frac{\phi(x_s)}{x_s}
-\phi^\prime(x_n)\right ]
\end{equation}
On the other hand, from eqn.(27) if we consider 
\begin{equation}
p_F=\frac{Ze^2}{\mu x}(\phi(x)^2-\phi_0^2x^2)^{1/2},
\end{equation}
then in the ultra-relativistic approximation we have the same result as 
shown in eqn.(32). The cell averaged electron kinetic energy density is 
then given by
\begin{equation}
\epsilon_{KE}=\frac{E_{KE}}{V}=\frac{3E_{KE}}{4\pi r_s^3}
=\frac{3E_{KE}}{4\pi\mu^3 x_s^3}
\end{equation}
The corresponding expression for $B=0$ in the non-relativistic regime is given 
by \cite{SHA}
\[
E_{KE}=\frac{3}{7}\frac{Z^2e^2}{\mu}\left [
\frac{4}{5}x_s^{1/2}\phi^{5/2}(x_s) -\phi^\prime(0)\right ]
\]

\section{Interaction Energy:Electron-Nucleus}
Next we consider the three possible types of interaction term. Let us first 
consider the electron-nucleus interaction term, given by
\begin{eqnarray}
E_{en}&=&-Ze^2\int_{r_n}^{r_s}d^3r \frac{n_e}{r}\nonumber \\
&=&-4\pi Z e^2\mu^2\int_{x_n}^{x_s} xdx n_e(x)
\end{eqnarray}
where the exact expression for electron density $n_e(x)$ is given by eqn.(7) 
and the Fermi momentum $p_F(x)$ is given by eqn.(27). Then we have after 
expressing $p_F(x)$ as a function of $\phi(x)$
\begin{equation}
E_{en}=-\frac{Z^2e^2}{\mu} 
\int_{x_n}^{x_s}
\sum_{\nu=0}^{\nu_{\rm{max}}}(2-\delta_{\nu 0})
dx (\phi(x)^2-\phi_0^2x^2)^{1/2}
\end{equation}
This integral can not be evaluated analytically. To obtain the exact
values for $E_{en}$, as before, we need $\phi(x)$ as a function of $x$ and
$\nu_{\rm{max}}$ as a function of $\phi(x)$ within the integration
range. To get a simple form, we consider the ultra-relativistic limit. Then 
we have
\begin{equation}
E_{en}=-\frac{Z^2e^2}{\mu} 
\int_{x_n}^{x_s}
\sum_{\nu=0}^{\nu_{\rm{max}}}(2-\delta_{\nu 0})
dx \phi(x)
\end{equation}
Using the approximate form of the Poisson's equation, given by
\begin{equation}
\phi^{\prime\prime}(x)\approx \sum_{\nu=0}^{\nu_{\rm{max}}}(2-\delta_{\nu 0})
\phi(x)
\end{equation}
we have (again assuming $\nu_{\rm{max}}(x_s)$ instead of a function of
$x$)
\begin{equation}
E_{en}=-\frac{Z^2e^2}{\mu} \int_{x_n}^{x_s}dx \phi^{\prime\prime}(x)
\end{equation}
We finally have, after integrating by parts and using the surface condition
for $\phi(x)$
\begin{equation}
E_{en}=-\frac{Z^2e^2}{\mu}\left [ \frac{\phi(x_s)}{x_s}-\phi(x_n)\right ]
\end{equation}
The sum of electron kinetic energy and the electron-nucleus interaction
energy is then given by
\begin{equation}
E_{KE}+E_{en}=
-\frac{Z^2e^2}{2\mu}\left [ \frac{\phi(x_s)}{x_s}-\phi(x_n)\right ]
\end{equation}
and the cell averaged electron-nucleus interaction energy density is given by
\begin{equation}
\epsilon_{en}=\frac{E_{en}}{V}=\frac{3E_{en}}{4\pi r_s^3}
=\frac{3E_{en}}{4\pi\mu^3 x_s^3}
\end{equation}
The corresponding zero field value in the non-relativistic regime is given by
\cite{SHA}
\begin{eqnarray}
E_{en}&=&-\frac{Z^2e^2}{\mu}\int_0^{x_s}\phi^{3/2}(x)x^{-1/2}dx\nonumber \\
&=& -\frac{Z^2e^2}{\mu}\int_0^{x_s}\phi^{\prime\prime}(x)dx\nonumber \\
&=&-\frac{Z^2e^2}{\mu}\left (\frac{\phi(x_s)}{x_s}-\phi^\prime(0)\right )\nonumber
\end{eqnarray}

\section{Interaction Energy:Electron-Electron Direct Interaction}
Next we consider the electron-electron interaction. Let us first
evaluate the direct term. It is given by
\begin{equation}
E_{ee}^{(d)}=\frac{1}{2}e^2\int d^3r n_e(r)\int d^3r^\prime n_e(r^\prime)
\frac{1}{\vert \vec r-\vec {r^\prime} \vert}
\end{equation}
Assuming $\vec r$ as the principal axis and $\theta$ is the angle between
$\vec r$ and $\vec {r^\prime}$, we have $d^3r=4\pi r^2 dr$, $d^3r^\prime= 2\pi
{r^\prime}^2dr^\prime \sin \theta d\theta$ (we have assumed that the vectors 
$\vec r$ and $\vec {r^\prime}$ are on the same plane) and $\vert \vec r-\vec
{r^\prime}\vert = (r^2+{r^\prime}^2-2rr^\prime \cos \theta)^{1/2}$. The limits
for both $r$ and  $r^\prime$ are from $r_n$ to $r_s$ and the range of 
$\theta$ is from $0$ to $\pi$. Let us first evaluate the angular integral, 
given by
\[
I(r,r^\prime)=\int_0^\pi \frac{\sin \theta}{
(r^2+{r^\prime}^2-2rr^\prime \cos \theta)^{1/2}}
\]
It is straight forward to show that $I(r,r^\prime)=(r+r^\prime)-\vert r -
r^\prime \vert$. Then we have
\begin{equation}
E_{ee}^{(d)}=4\pi^2e^2\int_{r_n}^{r_s}rdr n_e(r)\int_{r_n}^{r_s}r^\prime
dr^\prime n_e(r^\prime)[(r+r^\prime)-\vert r-r^\prime\vert]
\end{equation}
Now from this equation it is trivial to show that the quantity within the 
third bracket will be $2r^\prime$ for $r^\prime < r$ and for the opposite 
case it will be $2r$. Then the above expression, for direct interaction, 
reduces to the following simple form:
\begin{eqnarray}
E_{ee}^{(d)}&=&8e^2\pi^2 \Big \{\int_{r_n}^{r_s}rdr n_e(r)
\int_{r_n}^r{r^\prime}^2dr^\prime n_e(r^\prime)\nonumber \\
&+& 
\int_{r_n}^{r_s}r^2dr n_e(r)
\int_r^{r_s}r^\prime dr^\prime n_e(r^\prime)\Big \}
\end{eqnarray}
To obtain $E_{ee}^{(d)}$, we have evaluated separately the integrals 
appearing in the above expression. They are given by
\begin{eqnarray}
I_1&=&\int_{r_n}^{r_s}rdr n_e(r)\times I_2(r)=\mu^2
\int_{x_n}^{x_s}xdx n_e(x)\times I_2(x) \nonumber \\
I_3&=&\int_{r_n}^{r_s}r^2dr n_e(r)\times I_4(r)=\mu^3
\int_{x_n}^{x_s}x^2dx n_e(x)\times I_4(x) \nonumber 
\end{eqnarray}
where
\begin{eqnarray}
I_2(r)&=&\int_{r_n}^r{r^\prime}^2dr^\prime n_e(r^\prime)= \mu^3
\int_{x_n}^x{x^\prime}^2dx^\prime n_e(x^\prime)=I_2(x)\nonumber \\
I_4(r)&=&\int_r^{r_s}r^\prime dr^\prime n_e(r^\prime)=\mu^2
\int_x^{x_s}x^\prime dx^\prime n_e(x^\prime)=I_4(x)\nonumber
\end{eqnarray}
It is easy to verify that the only integral which can be evaluated 
analytically is $I_4(x)$. It is therefore necessary to obtain $I_1$, $I_2$  
and $I_3$ numerically. Therefore, the exact analytical express for
$E_{ee}^{(d)}$ can not be obtained. However, we can have some approximate 
analytical results. As for example, 
\begin{eqnarray}
I_2(x)&=&\mu^3 \int_{x_n}^x {x^\prime}^2 dx^\prime n_e(x^\prime)\nonumber \\
&=& \frac{e^3ZB\mu^2}{2\pi^2}
\int_{x_n}^x \sum_{\nu=0}^{\nu_{\rm{max}}}(2-\delta_{\nu 0})
(\phi^2-\phi_0^2{x^\prime}^2)^{1/2}x^\prime dx^\prime \nonumber \\
&\approx& \frac{e^3ZB\mu^2}{2\pi^2}
\int_{x_n}^x \sum_{\nu=0}^{\nu_{\rm{max}}}(2-\delta_{\nu 0})
\phi(x^\prime) x^\prime dx^\prime\nonumber \\
&\approx& \frac{e^3ZB\mu^2}{2\pi^2} \int_{x_n}^x
\phi^{\prime\prime}(x^\prime) x^\prime dx^\prime \nonumber \\
&=& \frac{Z}{4\pi}\left [(x\phi^\prime(x)-x_n\phi^\prime(x_n))-(\phi(x)-
\phi(x_n)) \right ]
\end{eqnarray}
Similarly, we have approximately
\begin{eqnarray}
I_1&\approx& \mu^2\int_{x_n}^{x_s} xdx n_e(x) 
\frac{Z}{4\pi}\left [(x\phi^\prime(x)-x_n\phi^\prime(x_n))-(\phi(x)-\phi(x_n))
\right ] \nonumber \\
&\approx& \frac{Z^2e^2eB\mu\alpha}{8\pi^3} \int_{x_n}^{x_s} \phi(x)dx
[(x\phi^\prime(x)-x_n\phi^\prime(x_n))-(\phi(x)-\phi(x_n))]
\end{eqnarray}
To evaluate this integral, we split it into four integrals, given by
\begin{eqnarray}
&& I_1^{(1)}=\int_{x_n}^{x_s}\phi(x)\phi^\prime(x) xdx \nonumber \\
&& I_1^{(2)}=I_1^{(4)}=\int_{x_n}^{x_s} \phi(x)dx\nonumber \\
&& I_1^{(3)}=\int_{x_n}^{x_s} (\phi(x))^2 dx\nonumber
\end{eqnarray}
The approximate results for $I_1^{(2)}$ (=$I_1^{(4)})$ is straight
forward and is given by
\[
I_1^{(2)}=I_1^{(4)}=\phi^\prime(x_s)-\phi^\prime(x_n)
\]
Similarly, 
\[
I_1^{(1)}=\frac{1}{2}\left [x_s\phi(x_s)^2-x_n\phi(x_n)^2\right ]i
~~~ {\rm{and}} ~~~ I_1^{(3)}=0
\]
Then we have approximately (again assuming that $\nu_{\rm{max}}$ is a
function of $x_s$)
\begin{eqnarray}
I_1 &\approx& \frac{Z^2e^2\mu eB}{8\pi^3}\Big [ \frac{\alpha}{2} \left \{
x_s\phi(x_s)^2-x_n\phi(x_n)^2\right \} \nonumber \\
&-& \phi^\prime(x_n)x_n\left (\phi^\prime(x_s)-\phi^\prime(x_n)\right ) +
\phi(x_n) \left ( \phi^\prime(x_s)-\phi^\prime(x_n)\right ) \Big ]
\end{eqnarray}
On putting this result in eqn.(45), the approximate form of direct part of 
electron-electron interaction energy contribution from the integral $I_1$ is 
given by
\begin{eqnarray}
E_{ee}^{(d)(1)}&\approx &
\frac{Z^2e^2}{2\mu}\Big [ \frac{\alpha}{2} \left \{
x_s\phi(x_s)^2-x_n\phi(x_n)^2\right \} \nonumber \\
&-& \phi^\prime(x_n)x_n\left (\phi^\prime(x_s)-\phi^\prime(x_n)\right ) +
\phi(x_n) \left ( \phi^\prime(x_s)-\phi^\prime(x_n)\right ) \Big ]
\end{eqnarray}
where $\alpha=\nu_{\rm{max}}(\nu_{\rm{max}}+1)$.
Similarly the integral $I_4$ is approximately given by
\begin{eqnarray}
I_4&\approx& \frac{eBZe^2\mu}{2\pi^2}
\int_x^{x_s} \sum_{\nu=0}^{\nu_{\rm{max}}}(2-\delta_{\nu 0})
\phi(x^\prime)dx^\prime \nonumber \\
&\approx& \frac{eBZe^2\mu}{2\pi^2}\int_x^{x_s}
\phi^{\prime\prime}(x^\prime)dx^\prime\nonumber \\
&=& \frac{eBZe^2\mu}{2\pi^2}\int_x^{x_s}
(\phi^\prime(x_s)-\phi^\prime(x))\nonumber \\
&=&\frac{Z}{4\pi \mu}
(\phi^\prime(x_s)-\phi^\prime(x))
\end{eqnarray}
Similarly the integral $I_3$ may also be approximated by the integral
\begin{equation}
I_3 \approx  \frac{\mu eBZ^2e^2}{8\pi^3}
\int_{x_n}^{x_s} \sum_{\nu=0}^{\nu_{\rm{max}}}(2-\delta_{\nu 0})
\phi(x)xdx (\phi^\prime(x_s)-\phi^\prime(x))
\end{equation}
This integral may be broken into two parts
\begin{equation}
I_3^{(1)}= \int_{x_n}^{x^s} \sum_{\nu=0}^{\nu_{\rm{max}}}(2-\delta_{\nu 0}) 
\phi(x)xdx ~~~{\rm{and}}~~~ 
I_3^{(2)}= \int_{x_n}^{x^s} \sum_{\nu=0}^{\nu_{\rm{max}}}(2-\delta_{\nu 0}) 
\phi^\prime(x)\phi(x)xdx
\end{equation}
The first one is approximately given by
\begin{eqnarray}
I_3^{(1)}&\approx& \int_{x_n}^{x_s}\phi^{\prime\prime}(x)xdx\nonumber \\
&=& (x_s\phi^\prime(x_s)-x_n\phi^\prime(x_n))- (\phi(x_s)-\phi(x_n))
\end{eqnarray}
and the approximate value for the second one is
\begin{eqnarray}
I_3^{(2)}&\approx& \int_{x_n}^{x_s} \phi^\prime(x)\phi^{\prime\prime}(x)xdx
\nonumber \\
&=& x_s(\phi^\prime(x_s))^2-x_n(\phi^\prime(x_n))^2
\end{eqnarray}
Combining these two results, we have the approximate form of $I_3$ as given
below:
\begin{eqnarray}
I_3 \approx \frac{\mu eB Z^2e^2}{8\pi^3}
\Big [ &&\{ (x_s\phi^\prime(x_s)-x_n\phi^\prime(x_n)) -(\phi(x_s)-\phi(x_n))\}
- \nonumber \\
&&\{ x_s(\phi^\prime(x_s))^2-x_n(\phi^\prime(x_n))^2\} \Big ]
\end{eqnarray}
The approximate form of direct part of electron-electron interaction energy,
coming from the integral $I_3$ may then be obtained by substituting eqn.(55)
into eqn.(45) and is given by
\begin{eqnarray}
E_{ee}^{(d)(2)}
\approx \frac{ Z^2e^2}{2\mu}
& \Big [ &\{ (x_s\phi^\prime(x_s)-x_n\phi^\prime(x_n)) -(\phi(x_s)-\phi(x_n))\}
- \nonumber \\
& \{ & x_s(\phi^\prime(x_s))^2-x_n(\phi^\prime(x_n))^2\} \Big ]
\end{eqnarray}
Finally adding eqns.(49) and (56), we have the approximate form of direct 
part of electron-electron interaction energy
\begin{equation}
E_{ee}^{(d)}= E_{ee}^{(d)(1)}+ E_{ee}^{(d)(2)}
\end{equation}
The cell averaged direct interaction energy density is then given by
\begin{equation}
\epsilon_{ee}^{(d)}=\frac{E_{ee}^{(d)}}{V}=\frac{3E_{ee}^{(d)}}{4\pi r_s^3}
=\frac{3E_{ee}^{(d)}}{4\pi \mu^3 x_s^3}
\end{equation}
The corresponding direct part of e-e interaction term for the zero field case 
in the non-relativistic regime is given by \cite{SHA}
\[
E_{ee}^{(d)}=\frac{Z^2e^2}{2\mu} \int_0^{x_s}x^{1/2}\phi^{3/2}(x)dx\left \{
\frac{1}{x} \int_0^x{x^\prime}^{1/2}\phi^{3/2}(x^\prime)dx^\prime +
\int_x^{x_s}{x^\prime}^{-1/2}\phi^{3/2}(x^\prime)dx^\prime\right \}
\]
It is straight forward to evaluate these integrals using the Poisson's equation
and the surface condition. Then we have
\[
E_{ee}^{(d)}=\frac{1}{2}\frac{Z^2e^2}{\mu}\left [ -\frac{4}{7}x_s^{1/2}
\phi^{5/2}(x_s)-\frac{2}{7} \phi^\prime(0)\right ]
\]
The exact values for $E_{ee}^{(d)}$ can only be obtained numerically,
knowing $\phi(x)$ as a function of $x$ within the range $x_n$ to $x_s$.
This is true even for the field free case.

\section{Interaction Energy:Electron-Electron Exchange Term}
Next we shall consider the exchange term. The exchange energy integral 
corresponding to the $i$th. electron in the cell is given by
\begin{equation}
E_{ee}^{(ex)}= \frac{e^2}{2}\sum_j \int d^3r d^3r^\prime \frac{1}{\vert \vec
r-\vec {r^\prime} \vert} \bar\psi_i(\vec r)\bar\psi_j(\vec {r^\prime})
\psi_j(\vec r)\psi_i(\vec {r^\prime})
\end{equation}
where the spinor wave function is given by eqns.(2)-(4) and $\bar \psi(\vec r)=
\psi^\dagger(\vec r)\gamma_0$, the adjoint of the spinor and $\gamma_0$ is
the zeroth part of the Dirac gamma matrices $\gamma_\mu$. Now it is very
easy to show that for $t=t^\prime$
\begin{eqnarray}
\bar \psi_i(\vec r)\psi_i(\vec{r^\prime}) &=& \frac{2m}{L_yL_zE_\nu}
\exp[-i\{ p_y(y-y^\prime)+p_z(z-z^\prime)\}]\nonumber \\ &&
\{I_{\nu;p_y}(x)I_{\nu;p_y}(x^\prime)+I_{\nu-1;p_y}(x) I_{\nu-1;p_y}(x^\prime)\}
\end{eqnarray}
Similarly, we have
\begin{eqnarray}
\bar \psi_j(\vec {r^\prime})\psi_j(\vec r) &=& \frac{2m}{L_yL_zE_\nu^\prime}
\exp[i\{ p_y^\prime(y-y^\prime)+p_z^\prime(z-z^\prime)\}]\nonumber \\ &&
\{I_{\nu^\prime;p_y^\prime}(x)I_{\nu^\prime;p_y^\prime}(x^\prime)+
I_{\nu^\prime-1;p_y^\prime}(x) I_{\nu^\prime-1;p_y^\prime}(x^\prime)\}
\end{eqnarray}
When these two terms are combined, we have after replacing the sum over $j$
by the integrals
\[
L_yL_z\int_{-\infty}^{+\infty} dp_y^\prime
\int_{-p_F}^{+p_F} dp_z^\prime
\]
\begin{eqnarray}
E_{ee}^{(ex)}&=&\left (\frac{e^2}{2}\right )\left (
\frac{4m^2}{L_y^2L_z^2 E_\nu}\right )
\sum_{\nu^\prime=0}^{\nu_{\rm{max}}}(2-\delta_{\nu^\prime 0}) \int...\int
L_ydp_y^\prime L_zdp_z^\prime d^3rd^3r^\prime\frac{1}{E_{\nu^\prime}}
\frac{1} {\vert \vec r -\vec {r^\prime}\vert} \nonumber \\ &&
\exp[-i\{ (p_y-p_y^\prime)(y-y^\prime)+(p_z-p_z^\prime)(z-z^\prime)\}]
\nonumber \\ && [\{I_{\nu;p_y}(x)I_{\nu;p_y}(x^\prime)+I_{\nu-1;p_y}(x)
I_{\nu-1;p_y}(x^\prime)\} \nonumber \\
&& \{I_{\nu^\prime;p_y^\prime}(x)I_{\nu^\prime;p_y^\prime}(x^\prime)+
I_{\nu^\prime-1;p_y^\prime}(x) I_{\nu^\prime-1;p_y^\prime}(x^\prime)\}]
\end{eqnarray}
It is possible to evaluate the integrals over $y^\prime$ and $z^\prime$, given 
by \cite{LEE}
\begin{eqnarray}
\int_{-\infty}^{+\infty} \int_{-\infty}^{+\infty} dy^\prime dz^\prime 
\frac{1} {\vert \vec r -\vec {r^\prime}\vert} 
&& \exp[-i\{(p_y-p_{y^\prime})(y-y^\prime)+ (p_z-p_{z^\prime})(z-z^\prime) \} ]
\nonumber \\ &=& \frac{4\pi}{2K}\exp(-K\vert x-x^\prime\vert)
\end{eqnarray}
where $K=[(p_y-p_{y^\prime})^2+(p_z-p_{z^\prime})^2]^{1/2}$. Then the
integral over $y$ and $z$ is given by 
\[
\int_{-\infty}^{+\infty} \int_{-\infty}^{+\infty} dy dz=L_yL_z
\] 
Since the exchange integral is multi-dimensional in nature with an extremely 
complicated form of integrand, it is just impossible to evaluate the exchange 
energy analytically. However, a simple form can be achieved if all the 
electrons occupy only the zeroth Landau level (i.e., $\nu_{\rm{max}}=0$). 
This special case have been discussed later in this article. In the numerical 
evaluation of the above exchange energy integral, we have put $p_z=p_F$, the 
electron Fermi energy and for the sake of convenience we make the following 
substitution:
\[
X=x-\frac{p_y}{eB}, ~~~ X^\prime=x^\prime-\frac{p_y}{eB}, ~~~
P_y=p_y^\prime-p_y ~~{\rm{and}}~~ P_z=p_z^\prime-p_z
\]
Since $-\infty \leq x \leq +\infty$, $-\infty \leq x^\prime \leq +\infty$
and $-\infty \leq p_y^\prime \leq +\infty$, the limits of the new variables
will also remain same. Further, we have used $-p_F \leq p_z^\prime \leq +p_F$
and as consequence, we have $0 \leq P_z \leq 2p_F$. Although the ranges of 
the variables $x,y,x^\prime$ and $y^\prime$ are from $-r_s$ to $-r_n$ and 
then from $+r_n$ to $+r_s$, for the sake of convenience we have considered 
the limits from $-\infty$ to $+\infty$, assuming that the electron density 
vanishes inside the nucleus and also outside the cell surface.

The exact values for the exchange integrals are evaluated numerically
using Monte-Carlo technique to obtain electron-electron exchange
interaction part. In fig.(3) we have plotted the exchange 
energy (in MeV) for various values of electron Fermi momentum (in MeV). We 
have fitted numerically the exchange energy as a function of Fermi momentum 
and is given by
\[
E_{ee}{(ex)}=E_0^{(ex)}\exp(\alpha p_F)
\]
where the parameters $E_0^{(ex)}=0.043$ and $0.062$ in MeV and $\alpha=0.409$ 
and $0.42$ in MeV$^{-1}$ for $B=10^{14}$G and $10^{16}$G respectively. It has
been observed that the minimum value of $p_F$ for which $E_{ee}^{(ex)}$
is non-zero increases with the increase in $B$. The qualitative nature
of the curves are exactly identical. However, $E_{ee}^{(ex)}$ 
increases with $B$ for a given $p_F$.

We shall now consider the Thomas-Fermi-Dirac model for $\nu=0$. We assume here
that the strength of magnetic field is extremely strong, so that electrons 
occupy only the zeroth Landau level ($\nu_{\rm{max}}=0$). In this scenario, 
the mathematical derivations are much more easier than $\nu_{\rm{max}} \neq 0$ 
case. In this approximation
\begin{equation}
\varepsilon_F=(p_F^2+m^2)^{1/2} ~~~{\rm{and}}~~~ n_e=\frac{eB}{2\pi^2}p_F
\end{equation} 
Further, we define the electron chemical potential
\begin{equation}
\mu_e=\varepsilon_F(r)-eV(r)-m=\left [ \frac{4\pi^2n_e^2}{(eB)^2}+m^2 \right
]^{1/2}-eV(r) -m ={\rm{constant}}
\end{equation}
Which gives
\begin{equation} 
n_e = \frac{eB}{2\pi^2}[2m(eV(r)+\mu_e)]^{1/2} \left [ 1+\frac{(eV(r)+\mu_e)} 
{2m}\right ]^{1/2} 
\end{equation}
Substituting this expression in the Poisson's equation (eqn.(8)) and
discarding the nuclear contribution, we have
\begin{equation}
\nabla^2V = 4\pi e
\frac{eB}{2\pi^2}[2m(eV(r)+\mu_e)]^{1/2} \left [ 1+\frac{(eV(r)+\mu_e)}
{2m}\right ]^{1/2}
\end{equation} 
Following the same procedure as we did for $\nu\neq 0$, we get 
\begin{equation} 
\frac{d^2\phi}{dx^2}=(x\phi)^{1/2} \left [ 1 + \lambda
\frac{\phi(x)}{x}\right ]^{1/2}
\end{equation}
where 
\[
\lambda=\frac{Ze^2}{2m\mu}, ~~~ r=\mu x  ~~~{\rm{where}}~~~ \mu=\frac{Z^{1/5}
\pi^{2/5}} {2^{3/5}e^{4/5}B^{2/5}m^{1/5}}
\]
The initial and the surface conditions remain same as we have for
$\nu_{\rm{max}}\neq 0$ case. Further, the form of this equation is such 
that there is no singularity at the origin. Therefore, in this case we 
need not have to integrate from $r_n$ (we can assume safely that the nucleus 
is a point object), or do not have to use the numerical prescription as 
discussed in reference \cite{FEY}.

Almost the same algebraic procedure is followed to obtain Thomas-Fermi-Dirac 
equation for the relativistic case with $B=0$ \cite{REMO}. In this case
\[
\varepsilon=[{\vec p}^2+m^2] ~~~ {\rm{and}} ~~~ n_e= \frac{1}{3\pi^2}p_F^3
\]
Here, we have
\begin{equation}
\mu_e=\varepsilon_F(r)-eV(r)-m=[(3\pi^2n_e)^{2/3}+m^2]^{1/2}-eV(r)-m 
={\rm{constant}}
\end{equation} 
Which gives
\begin{equation}
n_e = \left [\frac{2m(\mu_e+eV(r))}{3\pi^2}\right ]^{3/2} 
\left [ 1+\frac{(eV(r)+\mu_e)}
{2m}\right ]^{3/2}
\end{equation} 
Then substitution of these expressions in the Poisson's equation, gives
\begin{equation}
\frac{d^2\phi}{dx^2}=\frac{\phi^{3/2}}{x^{1/2}} \left [ 1+ \frac{Z}{Z_{cr}}
\frac{\phi}{x}\right ]^{3/2} - ~~{\rm{nuclear~~ contribution}}~~~
\end{equation}
where
\[
Z_{cr}=\left (\frac{3\pi}{4e^3}\right )^{1/2} ~~~{\rm{and}}~~~ \mu=
\frac{(3\pi)^{2/3}}{me^22^{7/3}Z^{1/3}}
\]
The initial and surface conditions are again same as for $B \neq  0$ and 
$\nu_{\rm{max}} \neq 0$. Unfortunately, it has singular nature at the origin. 
Which is removed artificially by taking the lower limit of $r$ or 
$x$-integration as $r_n$ or $x_n$ respectively or following reference 
\cite{FEY} (see \cite{REMO}). Similarly for the non-relativistic regime with 
$B\neq 0$ but $\nu_{\rm{max}}=0$, it is very easy to formulate all the above 
equations. Here
\[
\varepsilon_F(r)=\frac{p_F^2(r)}{2m} ~~~ {\rm{and}} ~~~ n_e=\frac{eB}{2\pi^2}p_F
\]
Further,
\[
\mu_e= \frac{p_F^2(r)}{2m} -eV(r)= {\rm{constant}}~~~
\]
It gives
\[
n_e=\frac{eB}{2\pi^2}[2m(\mu_e+eV(r))]^{1/2}
\]
Which after substituting in the Poisson's equation, we have
\begin{equation}
\frac{d^2\phi}{dx^2}=x^{1/2}\phi^{1/2}
\end{equation}
This equation again has no singularity at the origin. Further, in this case
\[
\mu=\frac{\pi^{2/5}Z^{1/5}}{2^{2/5}e^{1/5}B^{2/5}(2m)^{1/5}}
\]
but the boundary conditions again remain as usual.

Let us now consider the non-relativistic case with $B\neq 0$ and
$\nu_{\rm{max}}\neq 0$. In this case the energy eigen value is given by
\[
E_\nu=\frac{p_z^2}{2m}+\left ( \nu +\frac{1}{2}\right )\frac{eB}{m} -eV(r)
\]
Putting $p_z=p_F$, we have
\[
\mu_e=\frac{p_F^2}{2m}+\left ( \nu +\frac{1}{2}\right )\frac{eB}{m}
-eV(r)={\rm{constant}}
\]
Hence,
\begin{eqnarray}
n_e&=&\frac{eB}{2\pi^2}\sum_\nu p_F \nonumber \\
&=& \frac{eB}{2\pi^2}\sum_\nu \{2m(\mu_e+eV(r))-(2\nu +1)eB\}^{1/2} \nonumber
\end{eqnarray}
Following the same procedure as we did for the relativistic case with
$\nu_{\rm{max}} \neq 0$, the Poisson's equation reduces to
\begin{equation}
\frac{d^2\phi}{dx^2}=x^{1/2}\sum_\nu (\phi(x)-x\phi_\nu)^{1/2}
\end{equation}
where
\[
\mu= \frac{Z^{1/5}\pi^{2/5}}{2^{2/5}e^{2/5}(eB)^{2/5}(2m)^{1/5}} ~~~
{\rm{and}} ~~~
\phi_\nu=\frac{\nu eB\mu}{mZe^2}
\]
In this particular scenario, by inspection one can realize that the 
necessary condition to be satisfied to have physically acceptable solutions 
is that the quantity within the square root on the right hand side of eqn.(73) 
should not be negative, i.e.,
\[
\phi -x\phi_\nu \geq 0  ~~~ {\rm{which ~~gives}} ~~~
\nu_{\rm{max}} \leq \frac{mZe^2\phi(x)}{eB\mu x} -\frac{1}{2}
\]
The above form of the Poisson's equation with both $B$ and $\nu_{\rm{max}}\neq 0$, 
has no singularity at the origin. The boundary conditions are same as for the
relativistic case. The above condition imposed on the upper limit
of $\nu$ has to be checked at every steps of numerical integration. Since
$\nu_{\rm{max}} \geq 0$, we have
\[
\frac{mZe^2\phi(x)}{eB\mu x} \geq \frac{1}{2}
\]
The wave function $\psi(\vec r) \propto I_{\nu;p_y}(x)$ for the
electrons with $I_{\nu;p_y}$, given by eqn.(5), is quite complicated.
The primary reason is the presence of Hermite polynomial of non-zero order. 
The exchange energy in this case also can not be evaluated analytically. 
The expression is as complicated as we have derived for the relativistic 
case.

On the other hand, with zero magnetic field for non-relativistic scenario, 
the exchange energy is given by \cite{YSL} (see also \cite{MER})
\begin{equation}
E_{ee}^{(ex)}(p_z)=\frac{e^2}{2\pi}\left [ \frac{(p_F^2-p_z^2)}{p_z} 
\ln\left \vert
\frac{p_F+p_z}{p_F-p_z} \right \vert +2p_F\right ]
\end{equation}
Substituting $p_z=p_F$, we have
\begin{equation}
E_{ee}^{(ex)}=\frac{e^2}{\pi}p_F
\end{equation}

For the sake of completeness we shall now briefly discuss the evaluation of 
exchange energy for both the non-relativistic and relativistic cases in 
presence of ultra-strong magnetic field, i.e., $B\neq 0$ and $\nu_{\rm{max}} 
=0$. For both these cases, semi-analytic expressions can be obtained. The 
exchange energy in the non-relativistic regime is given by
\begin{equation}
E_{ee}^{(ex)}=\frac{e^2}{2} \sum_{j=1}^Z \int d^3rd^3r^\prime{\frac{1}{\mid 
\vec r -\vec r^\prime\mid}} \psi_i^*(\vec r) \psi_j(\vec r) \psi_j^*
(\vec r^\prime) \psi_i(\vec r^\prime) 
\end{equation}
For the zeroth Landau level, since $H_0(x)=1$, the wave function $\psi(r)$ for 
electron is given by
\begin{equation}
\psi(\vec r)= \frac {1}{(L_yL_z)^{1/2}}\left ( \frac{eB}{\pi}\right
)^{1/4} \exp\left [-\frac{eB}{2}\left (x-\frac{p_y}{eB}\right )^2 \right ] 
\exp[i(p_yy +p_zz)]
\end{equation}
Here also we can replace the sum over $j$ by the integrals $\int 
\int L_yL_zdp_y^\prime dp_z^\prime$. Writing $d^3r^\prime= dx^\prime 
dy^\prime dz^\prime$.  and following Lee \cite{LEE}, we have
\begin{eqnarray}
\int dy^\prime dz^\prime \frac{1}{\mid \vec r -\vec r^\prime \mid}
\exp[-i(p_y-p_y^\prime)(y&-&y^\prime) -i(p_z-p_z^\prime)(z-z^\prime)]
 \nonumber \\&=&\frac{4\pi}{2K} \exp\left (-K\mid x-x^\prime \mid \right )
\end{eqnarray}
where $K=[(p_y-p_y^\prime)^2+(p_z-p_z^\prime)^2]^{1/2}$.

Similarly for $d^3r=dxdydz$, the integral $\int dy dz=L_yL_z$. Then we have
\begin{eqnarray}
E_{ee}^{(ex)}&=& \frac{1}{2}\left ( \frac{eB}{\pi} \right ) 4\pi e^2 \int
dp_y^\prime dp_z^\prime dx dx^\prime \frac{1}{2K} \exp \left (-K\mid
x-x^\prime\mid \right )   \nonumber \\ && \exp\left [-\frac{eB}{2} \left \{
\left ( x-\frac{p_y}{eB} \right )^2+
\left ( x-\frac{p_y^\prime}{eB} \right )^2+
\left ( x^\prime-\frac{p_y}{eB} \right )^2+
\left ( x^\prime-\frac{p_y^\prime}{eB} \right )^2
\right \} \right ]
\end{eqnarray}

To evaluate the integrals over $x$ and $x^\prime$, we change the
integration variables to $X$ and $Y$, where $X=x-x^\prime$ and $Y=
(x+x^\prime)/2$.

Now 
\begin{equation}
\int_{-\infty}^\infty dX \exp\left (-K\mid X\mid \right ) 
\exp\left (-\frac{eB}{2}X^2\right ) =
\left (\frac{2\pi}{eB}\right )^{1/2} \exp \left (\frac{K^2}{2eB}\right ) 
{\rm{erfc}}\left(
\frac{K}{(2eB)^{1/2}}\right )
\end{equation}
where ${\rm{erfc}}(x)$ is the complimentary error function.

Then  we have
\begin{eqnarray}
E_{ee}^{(ex)}&=&e^2B \int \frac{1}{K} dp_y^\prime dp_z^\prime dY \left (\frac
{2\pi}{eB}\right )^{1/2} \exp\left (\frac{K^2}{2eB}\right ) {\rm{erfc}}\left 
(\frac{K} {(2eB)^{1/2}}\right ) \nonumber \\ && \exp\left [ -\frac{eB}{2} 
\left ( 4Y^2 + \frac{2p_y^2}{e^2B^2} + \frac{2p_y^{\prime 2}}{e^2B^2} -
\frac{4p_yY}{eB}- \frac{4p_y^\prime Y}{eB} \right )\right ]
\end{eqnarray}
The $Y$ integral is given by
\begin{equation}
\int_{-\infty}^\infty dY \exp \left [ -\frac{eB}{2} \left ( 2Y
-\frac{p_y+p_y^\prime}{eB}\right )^2 \right ]=\left (\frac{\pi}{2eB}\right
)^{1/2}
\end{equation}
Then we have after changing the integration variables from $p_y^\prime$ and
$p_z^\prime$ to $P_y=p_y-p_y^\prime$ and $P_z=p_z-p_z^\prime$
\begin{equation}
E_{ee}^{(ex)}=e^2\pi\int dP_y dP_z \frac{1}{(P_y^2+P_z^2)^{1/2}} \exp \left (\frac
{P_z^2}{2eB} \right ) {\rm{erfc}}\left [ \left (\frac{P_y^2+P_z^2}
{2eB}\right )^{1/2}
\right ]
\end{equation}
where the limit of $P_y$ is from $-\infty$ to $\infty$ and $P_z$ is from
$0$ to $2p_F$ for $p_z=p_F$.

Again putting $P_y=P_z\tan\theta$, we have 
\begin{equation}
E_{ee}^{(ex)}= e^2 \pi \int_0^{2p_F} dP_z \int_0^{\pi/2} \sec\theta ~~d\theta
~~{\rm{erfc}}\left ( \frac{\mid P_z\mid}{(2eB)^{1/2}} \sec\theta \right )
\exp \left (\frac{P_z^2}{2eB}\right )
\end{equation}
This is the form of semi-analytic expression for the exchange energy with
$\nu_{\rm{max}}=0$ in the non-relativistic regime. Further simplification of 
this expression is not possible. Therefore, these double integrals have been 
evaluated numerically as a function of  Fermi momentum $p_F$. The fitted 
functional form of $E_{ee}^{(ex)}$ is given by
\begin{equation}
E_{ee}^{(ex)}=\alpha [ 1- \exp(-\beta p_F)] 
\end{equation}
where the parameters $\alpha$ and $\beta$ vary with magnetic field
strength $B$ and are shown in the following table.

\noindent {\bf{Table-I}}
\bigskip

\begin{tabular} {|l|c|c|r|} \hline 
B (Gauss) & $10^{14}$ & $10^{15}$& $10^{17}$ \\ \hline
$\alpha$ (MeV) &$0.568$& $1.796$& $17.909$\\ \hline
$\beta$ MeV$^{-1}$&$3.412$&$1.067$&$0.109$\\ \hline
$\gamma$ &$0.506$&$0.527$&$0.658$\\ \hline
$C$&$0.973$&$0.870$&$0.386$\\ \hline
$x_s$&$3.096$&$3.170$&$4.404$\\ \hline
$r_s$ (\AA)&$0.402$ &$0.203$& $0.123$ \\ \hline
$v_s$& $-0.938556$& $-0.937365$& $-0.936123$ \\ \hline
$\phi_0$& $1.633$&$1.651$&$1.944$\\ \hline
$\xi$& $2.097$&$2.071$&$1.755$\\ \hline
$x_0$&$0.213$&$0.204$&$0.031$\\ \hline
$\rho$ (gm/cc)&$72.79$&$572.29$&$962.14$\\ \hline
\end{tabular}

\bigskip
Unlike the relativistic case here one can see from the fitted functional
form, that the exchange energy saturates to some constant value $\alpha(B)$. 
Now, if we include the exchange part separately, then in Thomas-Fermi-Dirac 
model the electron Fermi energy is given by,
\begin{equation}
\mu=\frac{p_F^2}{2m}-e\phi-E_{ee}^{(ex)}(p_F)={\rm{constant}}
\end{equation}
Rearranging the above equation in the form (see also \cite{fus}),
\begin{equation}
\frac{p_F^2}{2m}+\alpha e^{-\beta p_F} =\mu^*+e\phi
\end{equation}
where $\mu^*=\mu+\alpha$ is the modified form of Fermi energy of the
electron, one can express Fermi momentum $p_F$ as a function of $\mu^*+e\phi$. 
The numerically fitted functional form is given by a simple power law,
\begin{equation}
p_F=C(\mu^*+e\phi)^\gamma
\end{equation}
where, $C$ and $\gamma$  are constant parameters for a given magnetic field
strength. In Table-I above, we have shown the variation of $C$ and $\gamma$ 
with the magnetic field strength $B$. The variation of $\phi(x)$ with $x$ for a 
given magnetic field strength is given by the numerically fitted functional form
(the solution of the Poisson's equation is fitted numerically)
\begin{equation}
\phi(x)=\frac{\phi_0}{1+\exp\{\xi(x-x_0)\}}
\end{equation}
where, $\phi_0$, $\xi$, $x_0$  are constant parameters for a given magnetic
field strength. The variation of these parameters with magnetic field
strength are also shown in Table-I. In presence of strong quantizing magnetic 
field, the variation of $\phi$ with $x$ is entirely different from the 
non-magnetic case. The variation is more or less like the radial distribution 
of matter in neutron stars. Further, we use
\[
\frac{d\phi}{dx}\Big \vert_{x=x_s}=\frac{\phi(x_s)}{x_s}=v_s,
\]
We have also shown the variation of $v_s$ with $B$ in Table-I.

We next consider the relativistic case with $\nu_{\rm{max}}=0$. Since for 
$\nu=0$, we have $H_\nu(x)=1$ and $H_{\nu-1}(x)=0$, the appropriate form of 
exchange energy is given by
\begin{eqnarray}
E_{ee}^{(ex)}&=& e^2\int ...\int dp_y^\prime dp_z^\prime dx dx^\prime
\frac{4m^2}{E_FE_0^\prime} \frac{4\pi}{K} \exp(-K\vert x -x^\prime \vert)
\nonumber \\ &&
I_{0;p_y}(x)I_{0;p_y}(x^\prime)I_{0;p_y^\prime}(x)I_{0;p_y^\prime}(x^\prime)
\end{eqnarray}
where
\begin{eqnarray}
I_{0;p_y}(x)&=&\left ( \frac{eB}{\pi}\right )^{1/4} \exp\left (-\frac{1}{2}
eBX^2\right ) \nonumber \\ &&\left (X=x-\frac{p_y}{eB}\right )\nonumber \\
I_{0;p_y}(x^\prime)&=&\left ( \frac{eB}{\pi}\right )^{1/4} \exp
\left (-\frac{1}{2} eB{X^\prime}^2\right ) \nonumber \\
&&\left (X^\prime=x^\prime-\frac{p_y}{eB}\right )\nonumber \\
I_{0;p_y^\prime}(x)&=&\left ( \frac{eB}{\pi}\right )^{1/4} 
\exp\left [-\frac{1}{2} eB\left (X-\frac{P_y}{eB}\right )^2\right ] \nonumber \\
I_{0;p_y^\prime}(x^\prime)&=&\left ( \frac{eB}{\pi}\right )^{1/4} 
\exp\left [-\frac{1}{2} eB\left (X^\prime -\frac{P_y}{eB}\right )^2\right ] 
\nonumber \\ P_y&=&p_y^\prime-p_y \nonumber \\
P_z&=&p_z^\prime-p_z \nonumber \\ p_z&=&p_F \nonumber \\
E_0&=&({p_z^\prime}^2+m^2)^{1/2}\nonumber
\end{eqnarray}
We again get a similar type of semi-analytic expressions as shown in
eqn.(84). The physical quantities in this scenario are when evaluated 
numerically, can be fitted by the same type of functional forms, as shown by 
eqns.(85) and (87)-(89). The numerical values for the parameters are more or
less same as shown in Table-I. The qualitative nature of the
dependence of the parameters on the magnetic field remain almost unchanged. A 
complete numerical analysis of this formalism, along with the numerical 
estimate of equation of state of the crustal matter of 
strongly magnetized neutron stars will be presented in a future
communication. In that correspondence we shall also make
comparative studies of various models and approximations. In this
article we have presented very briefly some of the numerical estimates
to give a feeling of our formalism.

\section{Thomas-Fermi Induced Charge Density}
Finally we shall discuss the appearance of Thomas-Fermi induced charge
density inside the cells. The total charge density within the system is given by
\begin{equation}
\rho(r)=\rho_{\rm{ext}}(r)+\rho_{\rm{ind}}(r)
\end{equation}
where $\rho_{\rm{ind}}(r)$ is the induced charge density and is given by the 
fundamental equation of non-linear Thomas-Fermi theory: 
\begin{eqnarray}
\rho_{\rm{ind}}&=&-e[n_e(\mu_e+eV(r))-n_e(\mu_e)] \nonumber \\
&=& -e\frac{\partial n_e}{\partial \mu_e}\Big \vert_{V(r)=0}eV(r) \nonumber \\
&=& -e \frac{eB}{2\pi^2}
\sum_{\nu=0}^{\nu_{\rm{max}}}(2-\delta_{\nu 0})
\frac{E_F}{(E_F^2-m_\nu^2)^{1/2}} eV(r)\nonumber \\
&=& -e\frac{eB}{2\pi^2} \frac{1}{p_F}
\sum_{\nu=0}^{\nu_{\rm{max}}}(2-\delta_{\nu 0})(p_F^2+m_\nu^2)^{1/2}eV(r)
\end{eqnarray}
In the present case, we assume that the electric field $V(r)$ is a slowly
varying function of coordinate $r$. In actual practise, to obtain the charge
density, one has to solve self-consistently the Dirac equation in presence 
of an external magnetic field (in this case the strength $B>B^{(c)(e)}$) and the
electrostatic field $V(r)$ and the Poisson's equation satisfied by $V(r)$. In 
the exact scenario, the definition of electron density is given by
\begin{equation}
n_e(r)=\psi^\dagger(r) \psi(r)
\end{equation}
Hence one can obtain the total charge density and $V(r)$. However, the 
method is extremely complicated, even numerically. In Thomas-Fermi approach, 
we actually do not solve the Dirac equation, instead, assume that $V(r)$ is 
changing slowly with $r$ and get an approximate result. 

Now taking the Fourier transform of the last relation as given in eqn.(92), 
we get
\begin{equation}
\rho_{\rm{ind}}(q)=\chi(q)V(q)
\end{equation}
Then the Thomas-Fermi dielectric constant is given by
\begin{eqnarray}
\epsilon(q)&=&1-\frac{4\pi}{q^2}\chi(q)\nonumber \\ &=& 1+\frac{k_0^2}{q^2}
\end{eqnarray}
It is obvious that $\chi(q)$ (and hence $\epsilon(q)$) is independent of $q$. 
Now for the non-relativistic case ($m_\nu \gg p_F$)
\[
m_\nu\approx m+\frac{eB}{m}
\]
Then we have
\begin{eqnarray}
k_0^2&=&4\pi e^2 \frac{eB}{2\pi^2p_F}\left [
\sum_{\nu=0}^{\nu_{\rm{max}}}(2-\delta_{\nu 0})\left \{ m_\nu
+\frac{p_F^2}{2m_\nu}\right \}\right ]\nonumber \\
&=&4\pi e^2\frac{eB}{2\pi^2p_F}\left [ (2\nu_{\rm{max}}+1)m +\frac{eB}{m}
\nu_{\rm{max}}(\nu_{\rm{max}}+1) +
\sum_{\nu=0}^{\nu_{\rm{max}}}(2-\delta_{\nu 0})
\frac{p_F^2}{2m_\nu} \right]=k_{nr}^2
\end{eqnarray}
On the other hand, in the relativistic scenario ($p_F\gg m_\nu$), we have
\begin{equation}
\rho_{\rm{ind}}(r)=\left [-e^2\frac{eB}{2\pi^2}
\sum_{\nu=0}^{\nu_{\rm{max}}}(2-\delta_{\nu 0})\right ] V(r)
\end{equation}
Which gives
\begin{equation}
k_0^2=2e^2\frac{eB}{\pi}(1+2\nu_{\rm{max}})=k_{\rm{rel}}
\end{equation}
It can be shown that in presence of ultra-strong magnetic field for both the
non-relativistic and the relativistic cases
($\nu_{\rm{max}}$=0), 
\begin{equation}
k_{nr}^2=\frac{{k_{rel}}^2}{p_F}\left ( m+\frac{p_F^2}{2m}\right )
\end{equation}
To obtain the screened coulomb potential, we use the well known relation
\begin{equation}
V(q)=\frac{1}{\epsilon(q)}V_{\rm{ext}}(q)
\end{equation}
where 
\begin{equation}
V_{\rm{ext}}(r)=\frac{Q}{r}
\end{equation}
Hence, the screened coulomb potential is given by
\begin{equation}
V_{\rm{ind}}(r)=\frac{Q}{r} \exp(-k_0 r)
\end{equation}
Since $k_0$ is a function of magnetic field strength, the above equation
gives the screened coulomb potential in presence strong quantizing magnetic
field. In the numerical evaluation of the screening length in presence of
strong magnetic field, one can use any one of these expressions as given above.

In this context, we must mention that in a very recent work, Shabad and 
Usov have investigated the effect of strong magnetic field on coulomb 
potential \cite{USOV}. It has been shown that the coulomb potential gets 
modified significantly in presence of strong quantizing magnetic field. 
This is, unlike the Thomas-Fermi model, is an exact field theoretic approach. 
In this work the modified form of vacuum polarization (grossly speaking, 
this will give a modification of the screening length) in presence of strong 
quantizing magnetic field has been considered. We expect that a lot of new 
results can be obtained if one incorporates these results
in Thomas-Fermi model calculation for the crustal matter of
magnetars. In particular the interaction terms (electron-nucleus,
electron-electron direct and exchange terms) will be modified significantly and 
thereby affects the equation of states of this low density matter.
\section{Conclusions}
In this article we have developed the formalism for relativistic version
of Thomas-Fermi-Dirac model in presence of strong quantizing magnetic field.
The formalism is applicable to the outer crust of magnetars and also to
strongly magnetized white dwarfs.

We have compared our results with several other cases, e.g., the well known
non-relativistic model with zero magnetic field, field free relativistic
case, non-relativistic model in presence of strong quantizing magnetic field
for both $\nu_{\rm{max}}\neq 0$ and $\nu_{\rm{max}}=0$. 

We have noticed that in this formalism, to solve the Poisson equation numerically 
it is necessary to include a few more conditions, which were absent in the 
usual field free non-relativistic model or in presence of ultra-strong
magnetic field ($\nu_{\rm{max}}=0$).

To remove singularity at the origin, we suggest, following \cite{REMO}, to
use finite dimension for the nuclei. It has also been noticed that unlike 
other scenario, one extra condition appears in the non-relativistic regime 
with $B\neq 0$ and $\nu_{\rm{max}}\neq 0$.

We have also given an approximate method to get an estimate of the induced 
charge within each cell and thereby obtain the variation of screening length
with magnetic field strength.

In our model, the Wigner-Seitz cells are assumed to be spherical in nature
and found that the radius of each cell decreases with the increase of
magnetic field strength. The variation is given by $\sim B^{-1/2}$.

The formalism is of course not applicable to the inner crust region, where
the matter density is close to the neutron drip point, some of the neutrons may 
come out from the cells. In some future communication we shall present a
modified version of this formalism appropriate for the inner crust region.

We have assumed that all the electrons within the cells are moving freely, 
i.e., they are not bound in any one of the atomic orbitals. In reality, it
may happen that the electrons at the vicinity of the nucleus in a cell have
negative energy. These electrons, therefore can not be treated as free. It
is therefore absolutely necessary to get the total energy of an electron
as a function of its position ($r$ or $x$) within the cell from the numerical 
solution of the Poisson's equation and the expressions for kinetic and
various form of interaction energies. We expect that very close to the
nucleus, the electron energy will be negative and for a particular value of $x$
($=r/\mu$) (which may be a function of $B$) it will become zero (quasi-free 
electrons) and then becomes positive. If it is found so, then we can not assume
that all the $Z$-electrons in the cell are participating in statistical 
processes. On the other hand, if we consider the expression for electron 
energy as given in eqn.(86), then from the physics point of view all
the electrons will become free (energy is always positive). Whereas, if 
we consider 
\[
\mu=~~~{\rm{kinetic~ energy}}~~~ -e\phi=~~~{\rm{constant}},
\]
then we may have bound, quasi-free and free
electrons within the cells. The presence of free electrons in the
compressed cells in a dense medium is popularly known as 
{\it{statistical ionization}} (see reference \cite{SITE} for a detailed 
discussion).

To conclude our results, in the following we have given in tabular form the 
variation of Fermi momentum, Pressure, and various kinds of energy (except 
the exchange energy part) for electron gas within a typical Wigner-Seitz cell, 
with the strength of magnetic field.

\noindent {\bf{Table-II}}
\bigskip

\begin{tabular} {|c|c|c|c|c|c|} \hline\hline 
$B/B_c$ & $p_F(x_s)$(MeV) & $P(x_s)$(MeV$^4$) & $E_{KE}(x_s)$(MeV) & 
$E_{en}(x_s)$(MeV) & 
$E_{ee}^{(d)}(x_s)$(MeV) \\
\hline\hline
  $10^5$ &  $2.57$  &$8.64\times 10^4$&  $15.95$ & $46.71$&  $8.12$ \\
  $5\times 10^4$&  $2.58$&  $7153.02$&  $18.13$&  $27.15$&  $12.15$\\
$10^4$& $2.66$&  $4323.14$&  $19.58$&  $4.88$&  $19.80$\\
$5\times 10^3$&  $2.76$&  $681.79$&  $24.53$& $-0.37$&  $23.82$\\
$10^3$& $3.56$&  $319.78$&  $27.30$& $-7.89$&  $36.18$\\
$5\times 10^2$&  $4.82$&  $62.99$&  $37.67$& $-9.69$&  $44.55$\\
$10^2$& $6.95$&  $38.06$&  $44.81$& $-12.09$&  $78.04$\\
$50$&  $9.13$&  $31.49$&  $195.02$& $-13.13$&  $93.95$\\
$10$&  $13.61$&  $10.06$&  $781.47$& $-13.85$&  $104.24$\\
$1$&  $19.94$&  $6.28$&  $1.13\times 10^5$& $-35.94$&
$130.67$\\
\hline\hline
  \end{tabular}

From the above tabular form of data one can see that the electron Fermi 
momentum and the corresponding kinetic energy decreases with the strength of 
magnetic field. Since the exchange energy has to be subtracted and its magnitude
increases with the magnetic field strength, we may conclude that the 
system becomes more and more stable (total energy decreases) with the increase 
in magnetic field strength.

\newpage
\begin{figure}[ht]
\psfig{figure=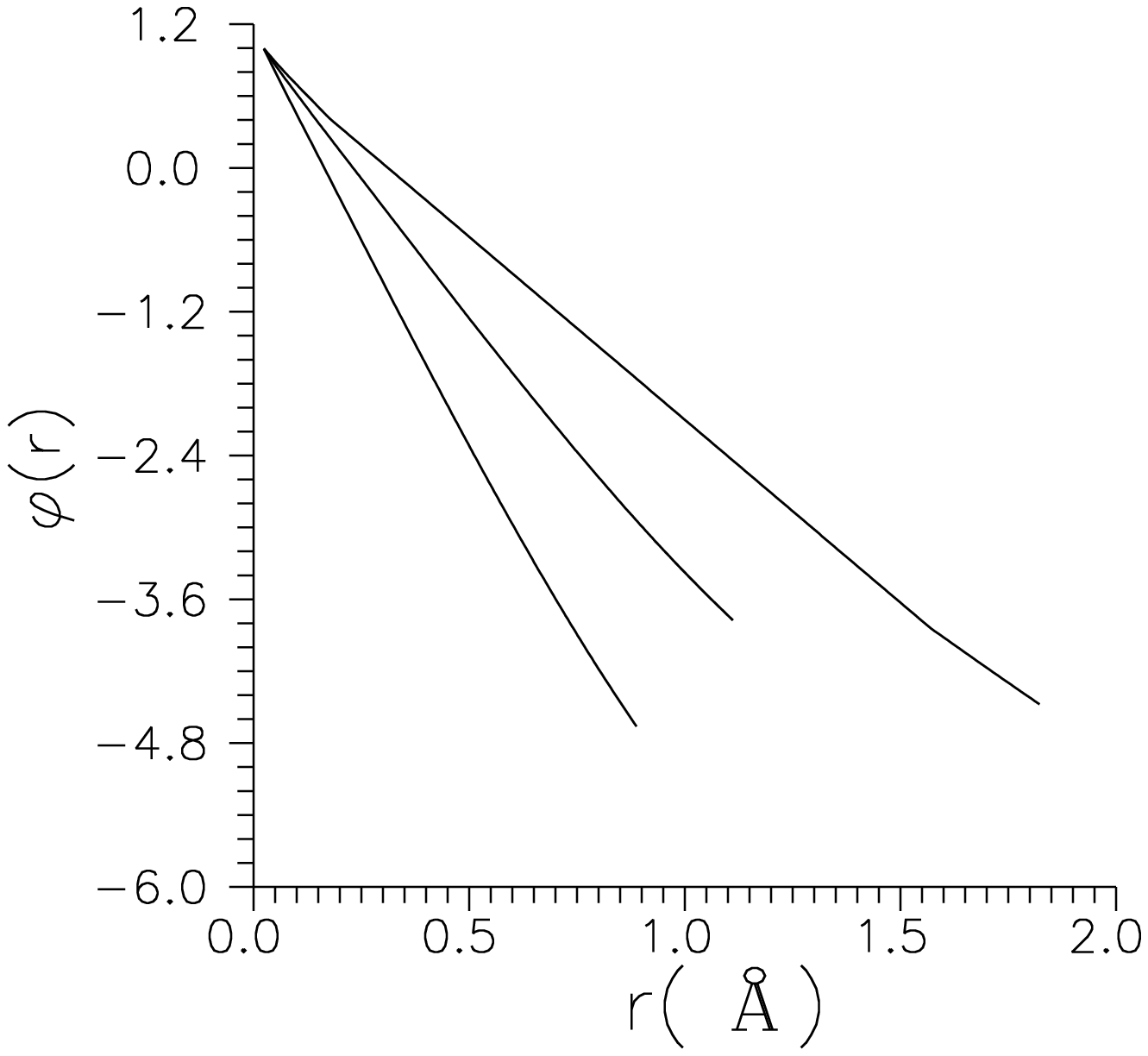,height=0.5\linewidth}
\caption{The variation of electrostatic field with radial distance from
the centre, for three different initial values:$\phi_{in}^\prime=-1.8$ (upper),
$\phi_{in}^\prime=-2.7$ (middle) and $\phi_{in}^\prime=-5.9$ (lower). The 
magnetic field strength $B=10^{14}$G}
\end{figure}
\begin{figure}[ht]
\psfig{figure=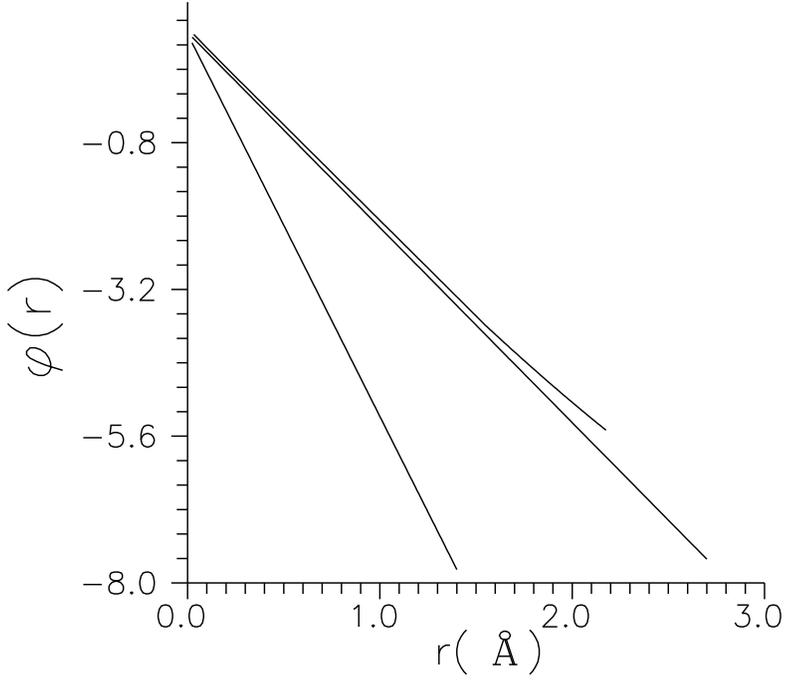,height=0.5\linewidth}
\caption{The variation of electrostatic field with radial distance from
the centre, for three different magnetic field strengths:$10^{14}$G
(upper), $10^{15}$G (middle) and $10^{17}$G (lower)}
\end{figure}
\begin{figure}[ht]
\psfig{figure=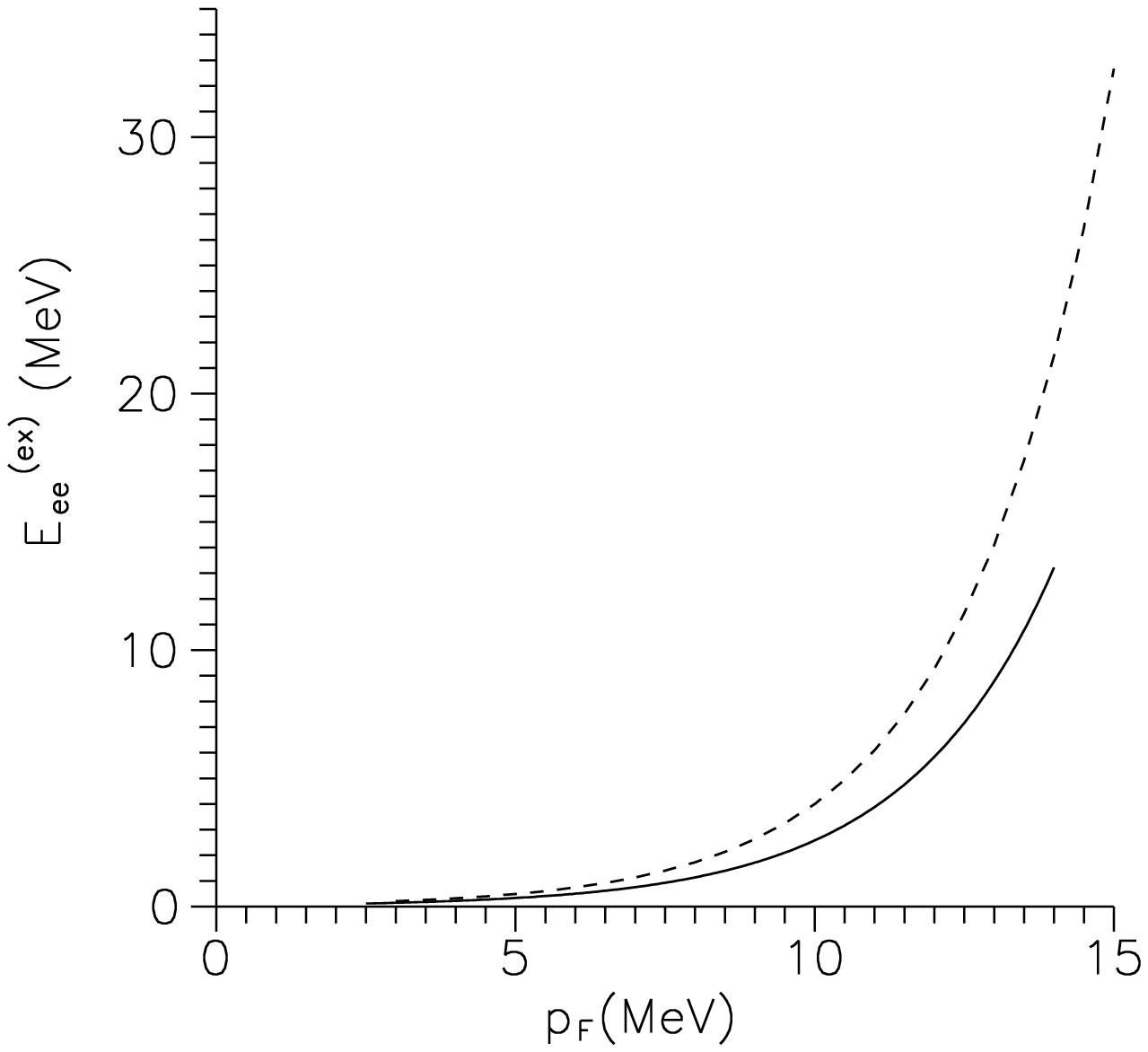,height=0.5\linewidth}
\caption{The variation of exchange energy with Fermi momentum of the
electrons. The magnetic field strength $B=10^{14}$G (solid curve) and
$10^{16}$G (dashed curve)}
\end{figure}

\begin{thebibliography}{99}
\bibitem{R1} R.C. Duncan and C. Thompson, Astrophys. J. Lett. {\bf{392}},
(1992), L9; C. Thompson and R.C. Duncan, Astrophys. J. {\bf{408}}, 
(1993), 194; C. Thompson and R.C. Duncan, MNRAS {\bf{275}}, (1995), 255;
C. Thompson and R.C. Duncan, Astrophys. J. {\bf{473}}, (1996), 322.
\bibitem{R2} P.M. Woods et. al., Astrophys. J. Lett. {\bf{519}},
(1999), L139; 
 C. Kouveliotou, et. al., Nature  {\bf{391}}, (1999), 235.
\bibitem{R3} K. Hurley, et. al., Astrophys. Jour. {\bf{442}}, (1999), L111.
\bibitem{R4} S. Mereghetti and L. Stella, Astrophys. Jour. {\bf{442}},
(1999), L17;
J. van Paradihs, R.E. Taam and E.P.J. van den Heuvel,
Astron. Astrophys. {\bf{299}}, (1995), L41; S. Mereghetti,
Invited review presented at the NATO ASI "The neutron star-black hole
connection", Elounda, Crete: 7-18 June, 1999; see also A. Reisenegger, 
Magnetic fields across the H-R diagram, ASP conference series,
{\bf{248}}, 469 (2001).
\bibitem{R5} D. Bandopadhyaya, S. Chakrabarty, P. Dey
and S. Pal, Phys. Rev. {\bf{D58}}, (1998), 121301.
\bibitem{R6} S. Chakrabarty, D. Bandopadhyay and S. Pal, Phys. Rev.
Lett. {\bf{78}}, (1997), 2898;
D. Bandopadhyay, S. Chakrabarty and S. Pal, Phys. Rev.
Lett. {\bf{79}}, (1997), 2176.
\bibitem{R7} C.Y. Cardall, M. Prakash and J.M. Lattimer,
astro-ph/0011148 and references therein;
E. Roulet, JHEP, {\bf{9801}}, 013 (1998); L.B. Leinson and A. P\'{e}rez, 
Astro-ph/9711216;
D.G. Yakovlev and A.D. Kaminkar, The Equation of States in
Astrophysics, eds. G. Chabrier and E. Schatzman P.214, Cambridge Univ.
\bibitem{R8}S. Chakrabarty and P.K. Sahu, Phys. Rev. {\bf{D53}}, (1996), 4687.
\bibitem{SCANN} S. Ghosh, S. Mandal and S. Chakrabarty, Ann. Phys.
{\bf{312}} (2004) 398.
\bibitem{SCLAND} S. Mandal, R. Saha, S. Ghosh and S. Chakrabarty, 
Phys. Rev. {\bf{C74}}, (2006), 015801.
\bibitem{R9}S. Chakrabarty, Phys. Rev. {\bf{D51}}, (1995), 4591;
Chakrabarty, Phys. Rev. {\bf{D54}}, (1996), 1306.
\bibitem{R10} T. Ghosh and S. Chakrabarty, Phys. Rev. {\bf{D63}},
(2001), 0403006; T. Ghosh and S. Chakrabarty, Int. Jour. Mod. Phys.
{\bf{D10}}, (2001), 89.
\bibitem{R11} V.G. Bezchastrov and P. Haensel, Phys. Rev. {\bf{D54}},
(1996), 3706.
\bibitem{R12} S. Mandal and S. Chakrabarty (in preparation).
\bibitem{RR13} A. Melatos, Astrophys. Jour. {\bf{519}}, (1999), L77;
A. Melatos, MNRAS {\bf{313}}, (2000), 217.
\bibitem{RR14} S. Bonazzola et al, Astron. \& Atrophysics. {\bf{278}},
(1993), 421; M. Bocquet et al, Astron. \& Atrophysics. {\bf{301}},
(1995), 757.
\bibitem{RR15} B. Bertotti and A.M. Anile, Astron. \& Atrophysics.
{\bf{28}}, (1973), 429; C. Cutler and D.I. Jones, Phys. Rev. {\bf{D63}},
(2001) 024002;
 K. Konno, T. Obata and Y. Kojima, Astron. \& Astrophys. {\bf{352}},
 (1999), 211; A.P.
Martinez et al, Phys. Rev. Lett. {\bf{84}}, (2000), 5261;
 M. Chaichian et al, Phys. Rev. Lett. {\bf{84}},
(2000), 5261; Guangjun Mao, Akira Iwamoto and Zhuxia Li, Chin. J.
Asstron. \& Astrophys. {\bf{3}}, (2003), 359;
A. Melatos, MNRAS {\bf{313}}, (2000), 217; R. Gonz\'{a}lez Felipe et al,
Chin. J. Astron. \& Astrophys. {\bf{5}}, (2005), 399 and references therein.
\bibitem{SOMA} S. Mandal and S. Chakrabarty, Int. Jour. Mod. Phys. {\bf{D13}} 
(2004) 1157.
\bibitem{RR16} S. Chakrabarty, Astrophysics \& Space Science  
{\bf{310}}, (2007) 195.
\bibitem{SCCH} S. Ghosh, S. Mandal and S. Chakrabarty, Phys. Rev. 
{\bf{C75}}, (2007) 015805.
\bibitem{R13} V.P. Gusynin, V.A. Miransky and I.A. Shovkovy, Phys. Rev.
{\bf{D52}}, (1995), 4747.
\bibitem{R14} D.M. Gitman, S.D. Odintsov and Yu.I. Shil'nov, Phys. Rev.
{\bf{D54}}, (1996), 2968.
\bibitem{R15} D.S. Lee, C.N. Leung and Y.J. Ng, Phys. Rev. {\bf{D55}},
(1997), 6504.
\bibitem{R16} V.P. Gusynin and I.A. Shovkovy, Phys. Rev. {\bf{D56}},
(1995), 5251.
\bibitem{R17} D.S. Lee, C.N. Leung and Y.J. Ng, Phys. Rev. {\bf{D57}},
(1998), 5224.
\bibitem{R18} V.P. Gusynin, V.A. Miransky and I.A. Shovkovy, Phys. Rev.
Lett. {\bf{83}}, (1999), 1291.
\bibitem{R19} E.V. Gorbar, Phys. Lett. {\bf{B491}}, (2000), 305.
\bibitem{Rina} T. Inagaki, T. Muta and S.D. Odintsov, Prog. Theo. Phys. 
suppl. {\bf{127}}, (1997), 93; T. Inagaki, S.D. Odintsov and Yu.I. 
Shil'nov, Int. J.  Mod. Phys. {\bf{A14}}, (1999), 481; 
T. Inagaki, D. Kimura and T. Murata, Prog. Theo. Phys. {\bf{111}}, (2004), 371.
\bibitem{VP2} V.P. Gusynin, V.A. Miransky and I.A. Shovkovy, Phys. Lett.
{\bf{B349}}, (1995), 477.
\bibitem{SPK} S.P. Klevansky, Rev. Mod. Phys. {\bf{64}}, (1992) 649.
\bibitem{WO} C.N. Leung and S.-Y. Wang, Ann. Phys. {\bf{322}}, (2007),
701; Nucl. Phys. {\bf{B747}}, (2006), 266.
\bibitem{wang} Shang-Yung Wang, Phys. Rev. {\bf{D77}}, (2008), 025031.
\bibitem{ADLER} S. L. Adler, J. N. Bahcall, C. G. Callan, and M. N. Rosenbluth,
Phys. Rev. Lett. {\bf {25}} (1970) 1061; S.L. Adler, Ann. Phys. {\bf{67}},
(1971) 599.
\bibitem{ADLER1} S.L. Adler and C. Schubert, Phys. Rev. Lett. {\bf{77}},
(1996) 1695.
\bibitem{WILKE} C. Wilke and G. Wunner, Phys. Rev. {\bf{D55}}, (1997) 997.
\bibitem{BAIER} V.N. Baier, A.I. Milstein and R. Zh. Shaisultanov, Phys.
Rev. Lett. {\bf{77}}, (1996) 1691.
\bibitem{RU} A.V. Potekhin, Astron. \& Astrophys. {\bf{306}}, (1996),
999; A.V. Potekhin and D.G. Yaklovlev, Astron. \& Astrophys. {\bf{314}},
{1996}, 341.
\bibitem{NAG} N. Nag and S. Chakrabarty, Int. Jour. Mod. Phys. {\bf{D11}} 
(2002), 817.
\bibitem{SHIV} B.K. Shivamoggi and P. Mulser, Euro. Phys. Lett. {\bf{22}},
(1993), 657.
\bibitem{REMO} R. Ruffini, "Exploring the Universe", a Festschrift in
honour of Riccardo Giacconi, Advance Series in Astrophysics and
Cosmology, World Scientific, Eds. H. Gursky, R. Rufini and L. Stella, Vol. 
{\bf{13}}, (2000), pp. 383; Int.Jour. of Mod. Phys. {\bf{5}}, 
(1996) 507.
\bibitem{LIEB} E.H. Lieb and B. Simon, Phys. Rev. Lett. {\bf{31}}, (1973), 681.
\bibitem{LIEB1} E.H. Lieb, J.P. Solovej  and J. Yngvason, Phys. Rev. Lett. 
{\bf{69}}, (1992), 749.
\bibitem{LIEB2} E.H. Lieb, Bull. Amer. Math. Soc., {\bf{22}}, (1990), 1.
\bibitem{K1} B.B. Kadomtsev, Sov. Phys. JETP {\bf{31}}, (1970), 945.
\bibitem{MU12} R.O. Mueller, A.R.P. Rau and L. Spruch, Phys. Rev. Lett.
{\bf{26}}, (1971), 1136; A.R.P. Rau, R.O. Mueller and L. Spruch, Phys. Rev. 
{\bf{A11}}, (1975), 1865. 
\bibitem{HILL} S.H. Hill, P.J. Grout and N.H. March, Jour. Phys. {\bf{B18}},
(1985), 4665.
\bibitem{FEY} R.P. Feynman, N. Metropolis and E.Teller, Phy. Rev.
{\bf{75}}, (1949), 1561.
\bibitem{SHA} S.L. Shapiro and S.A. Teukolsky, Black Holes, White Dwarfs
and Neutron Stars, John Wiley and Sons, New York, (1983).
\bibitem{MER} N.W. Aschroft and N.D. Mermin, Solid State Physics,
Saunders College Publishing, New York, (1976).
\bibitem{LEE} T.D. Lee, Elementary particles and the universe: Eassays in
honous of M. Gellmann, ed. John. H. Schwarz, Cambridge University
Press, (1991), pp 135.
\bibitem{YSL} Physics of Dense Matter, Y.S. Leung, World Scientific,
(1984), pp 34.
\bibitem{fus} I. Fushiki,  E.H. Gudmundsson,  C.J. Pethick and J. Yngvason
Ann. Phys  {\bf{216}}, (1992), 29.
\bibitem{USOV} A.E. Shabad and V.V. Usov, Phys. Rev. Lett., {\bf{98}},
(2007), 180403; Phys. Rev. {\bf{D77}}, (2008), 025001.
\bibitem{SITE} L. Delle Site, Physica {\bf{293}}, (2001), 71.
\end{thebibliography}
\end{document}